# Persistent and anti-persistent stride-to-stride fluctuations: an ARFIMA decomposition consistent with closed-loop sensorimotor control

Preprint — not peer-reviewed.

Philippe Terrier

*HE-ARC, HES-SO University of Applied Sciences and Arts Western Switzerland, Neuchâtel, Switzerland*

**Corresponding author:** Philippe Terrier

E-mail: Philippe.terrier@he-arc.ch

## Abstract

Stride-to-stride fluctuations in human walking carry a fractal correlation structure that reverses sign under external cueing: self-paced gait is persistent, whereas metronomic or visually cued gait is anti-persistent. Three decades of detrended fluctuation analysis (DFA) have established this reversal as a scaling-exponent shift, but DFA cannot distinguish genuine long-memory dynamics from short-memory autoregressive moving-average (ARMA) processes that produce the same apparent exponent. We fit the full eight-model ARFIMA(1,d,1) family to stride interval and stride speed series from three independent datasets (N = 70 subjects) spanning overground walking, fixed-speed treadmill walking, metronomic and visual cueing, and graded positional constraint. Model evidence is aggregated through BIC-based Schwarz weights, and the fractional differencing parameter d together with the autoregressive and moving-average coefficients φ and θ are estimated by Bayesian model averaging. Three findings emerge. (i) Long-memory specifications decisively outweigh ARMA alternatives under both persistent and anti-persistent conditions, establishing cued gait anti-persistence as a genuine fractional phenomenon. (ii) DFA α overestimates d + 0.5 by 0.25 to 0.34 units, a discrepancy jointly attributable to short-memory components that DFA conflates with long-memory persistence and to a finite-sample negative bias inherent to exact ML-ARFIMA estimation. (iii) The estimated (d, φ, θ) parameters are consistent with a corrective sensorimotor model in which a fractal intrinsic generator, a reactive feedback correction, and a motor-delay component together shape stride-to-stride fluctuations. Whether a single mechanistic model can account quantitatively for the observed parameter ranges across rhythmic, spatial, and unconstrained conditions is a question that the present analysis motivates but cannot alone resolve.

# 1. Introduction

The stride-to-stride fluctuations of human walking are not random noise [1]. During self-paced overground or treadmill locomotion, successive stride intervals (a stride being one complete gait cycle, comprising two steps from heel-strike to heel-strike of the same foot) show *persistent long-range correlations*: a stride that is slightly longer than average tends, on average, to be followed by further long strides, and this tendency decays as a power law that extends across dozens or even hundreds of subsequent strides [1–3]. This fractal temporal structure, quantified by the scaling exponent $\alpha \approx 0.7 - 1.0$ of detrended fluctuation analysis (DFA), has been interpreted as a signature of the hierarchical, multi-timescale organization of the locomotor control system: spinal, brainstem, and cortical circuits interact across overlapping timescales to produce scale-free variability [1,4,5]. The fractal signature is clinically informative. It degrades with aging, Parkinson's disease, and Huntington's disease, and its preservation has been associated with fall resistance and gait automaticity [5–7].

This hierarchical control architecture of walking governs three mechanically interdependent outputs: stride time (ST, the duration of one gait cycle), stride length (SL, the distance covered per cycle), and stride speed (SS = SL/ST). Because these three quantities are mechanically linked, the locomotor control system can satisfy any given speed by trading off ST against SL. This redundancy may be the source of fractal persistence under unconstrained walking conditions. Terrier, Turner, and Schutz were the first to demonstrate, using GPS-instrumented overground walking of long duration, that all three parameters simultaneously show persistent long-range correlations ($\alpha \approx 0.7 - 1.0$), extending the fractal dynamics framework beyond stride time alone [8].

When an external constraint is imposed on one of these parameters, fixing cadence, speed, or step length, the locomotor system can no longer exploit the ST–SL redundancy freely, and the fractal persistence of the constrained parameter gives way to anti-persistence. Hausdorff and colleagues first reported in 1996 that synchronizing heel strikes to an isochronous metronome replaces the persistent ST correlation structure with an anti-persistent one ($\alpha \approx 0.2 - 0.4$) [4,9]. Terrier et al. [8] later demonstrated that all three spatiotemporal parameters exhibit fractal persistence during unconstrained overground walking, and that under metronome pacing, SL and SS retain this structure, confining the anti-persistent shift to the temporally constrained parameter (ST). The selective disruption of ST alone reflects the specificity of the imposed constraint: the metronome targets cadence, so only the temporal parameter is tightly regulated; the spatial and speed parameters remain free to fluctuate fractally. A complementary dissociation was identified by Dingwell and Cusumano: treadmill walking without any metronome cueing renders SS anti-persistent ($\alpha \approx 0.2 - 0.4$) while leaving ST and SL persistent [10,11]. On a constant-speed belt, the walker must regulate stride speed to avoid drifting forward or backward; speed is now the constrained quantity, and the overcorrection dynamics generate anti-persistence exclusively in SS. The spatial parameter SL and the temporal parameter ST retain their redundancy and remain persistently correlated. Because ST, SL, and SS are mechanically coupled through SL = SS × ST, this pattern suggests what we refer to as the loss-of-redundancy hypothesis: persistent long-range correlations require that the locomotor system retain at least one mechanical degree of freedom within the ST–SL–SS manifold, and anti-persistence emerges in any parameter whose value is pinned by an external reference.

The hypothesis makes a sharper prediction when two constraints are imposed simultaneously: with both SS and ST pinned, SL is mechanically determined, no degree of freedom remains, and all three parameters should shift to anti-persistence. Terrier and Dériaz combined constant-speed treadmill walking with isochronous auditory cueing and found that *all three parameters*, ST, SL, and SS, became anti-persistent [12]. When the treadmill fixes SS and the metronome fixes ST, SL is fully determined (SL = SS × ST); the redundancy disappears, and the overcorrection dynamics propagate across the entire spatiotemporal parameter space. The effect was robust across three walking speeds and produced large effect sizes (Hedges' $g \approx 2$–4 for the shift from persistence to anti-persistence).

A remaining open question at that point was whether a *spatial* constraint, rather than a temporal one, could produce an equivalent transformation. Terrier addressed this in 2016 using a treadmill with augmented-reality projection capabilities (C-Mill, ForceLink), where stepping stones displayed on the belt imposed a spatial stepping target for each stride [13]. This visual cueing (VC) condition imposed a stride-length goal while the constant belt speed continued to impose a stride-speed goal. As predicted by the loss-of-redundancy model, VC produced anti-persistence in ST, SL, and SS, a pattern statistically indistinguishable from that observed under auditory cueing, despite engaging a fundamentally different sensory modality and targeting the spatial rather than temporal dimension of gait. In addition, the effect of cueing on gait *variability* (fluctuation magnitude, quantified by the coefficient of variation) diverges from its effect on correlation structure: auditory cueing leaves variability largely unchanged relative to unconstrained walking, whereas visual cueing substantially increases it (CV increase of 51–73% across parameters) [13], suggesting that spatial targeting imposes a higher attentional demand than temporal entrainment.

That the positional constraint, rather than any imposed cadence, drives SS anti-persistence is supported by converging evidence that this property is continuously tunable rather than categorical. Roerdink and colleagues projected walking areas of three different sizes onto the treadmill belt (small, intermediate, and the full 3-m belt), with and without isochronous acoustic pacing [14]. SS was anti-persistent in all six conditions (confirming that belt speed alone induces speed regulation) and the degree of anti-persistence scaled systematically with walking-area size, tightening as the area shrank. The same graded response appears at the opposite end of the constraint gradient. Choi and colleagues compared fixed-speed and self-paced treadmills [15]: on the fixed-speed belt SS was anti-persistent as expected, but on the self-paced belt (where the speed target tracks the walker rather than the reverse) SS reverted to persistence, indistinguishable from the overground baseline. Adding a metronome on the self-paced treadmill rendered ST anti-persistent while leaving SS persistent, exactly as in overground experiments [8]. A complementary manipulation by Decker, Cignetti, and Stergiou [16] showed that a concurrent executive task weakened SS anti-persistence during fixed-speed treadmill walking (α rising from 0.31 to 0.45), implicating conscious cortical resources in the correction. Taken together, these studies establish that SS anti-persistence is not a fixed state but a continuously tunable property of the speed-regulation dynamics: the tighter the positional constraint and the greater the attentional commitment, the more aggressively the system corrects stride-speed deviations, and the lower the DFA scaling exponent.

All of the evidence summarized so far rests on a single analytical method: detrended fluctuation analysis. A DFA scaling exponent $\alpha < 0.5$ is routinely interpreted as anti-persistent long-range dependence, and the graded pattern across studies has been read accordingly. This interpretation carries a strong and rarely examined assumption: the observed correlation structure requires a long-memory (fractal) component at all. A short-memory autoregressive moving-average (ARMA) process can produce the same sub-0.5 α without any fractional dynamics whatsoever. This distinction is not merely semantic. A simple first-order negative feedback loop, for example an MA(1) process with a negative moving-average coefficient $\theta$, produces a DFA exponent well below 0.5 even though it has no long-range temporal structure whatsoever [17,18]. In such a process, the autocorrelation is negative at lag 1 and exactly zero for all lags ≥ 2; the correction affects only the immediately adjacent stride. By contrast, a genuinely anti-persistent long-memory process, characterized by a negative fractional differencing parameter $d$ in the ARFIMA framework, shows autocovariances that decay hyperbolically across timescales [19,20].

The limitations of DFA extend beyond the long-memory question. DFA provides no information about the short-range ARMA structure that coexists with any long-memory component. The autoregressive parameter φ and moving-average parameter $\theta$, which capture the immediate error-correction dynamics of the sensorimotor control loop, are entirely invisible to DFA. DFA scaling exponents are also sensitive to the presence of ARMA components even in pure long-memory series: a positive autoregressive AR coefficient inflates the apparent DFA $\alpha$, while a negative moving-average MA coefficient deflates it, the net bias depends on the relative strength of each component [9,17,21]. These shortcomings motivate a parametric approach that can jointly estimate short-memory and long-memory components and formally test whether long-memory is present at all.

The ARFIMA(p,d,q) model family, introduced by Granger and Joyeux [19] and Hosking [20] and extensively developed in econometrics [21] provides this capability. Fitting ARFIMA and comparing it against nested ARMA alternatives using information criteria yields a direct statistical test of whether the observed autocorrelation structure requires a fractional differencing component. Torre, Delignières, and Lemoine introduced this approach to the human movement sciences in 2007 [17], using Akaike and Bayesian information criteria to compare 18 ARMA and ARFIMA models fitted to tapping and walking series. Wagenmakers and colleagues had earlier proposed a similar framework for cognitive serial correlations [22,23]. Delignières and Torre [9] applied the procedure to the original Hausdorff et al. [4]

gait dataset, finding negative *d* estimates under metronomic conditions and providing the first parametric evidence for anti-persistent long-range dependence in stride intervals. Subsequent ARFIMA applications in movement sciences have remained confined to tapping, finger coordination, and EEG signals, and have consistently used ARFIMA as a confirmatory diagnostic applied after DFA rather than as a primary model-selection framework [24–27].

The present paper addresses this gap by bringing the full ARFIMA model family to bear on three independent gait datasets (Hausdorff 1996 [4], Terrier 2016 [13], Roerdink 2019 [14]). Stride time is analyzed in the two datasets in which it was manipulated by metronomic cueing (Hausdorff, Terrier); stride speed is analyzed in the two datasets in which it was manipulated by walking-area width (Roerdink) or cueing modality (Terrier).

To interpret the parameter estimates produced by this approach, we draw on the sensorimotor synchronization literature. The persistent-to-anti-persistent transition has been studied in detail in finger tapping, the canonical paradigm of sensorimotor synchronization, where the structural ingredients are identical to metronomic walking: an isochronous external reference, an asynchrony between produced events and metronome beats, and a corrective adjustment of subsequent intervals. Our interpretive claim draws on two complementary contributions from this literature. Vorberg and Schulze [28] formalized the linear phase-correction framework: assuming a white-noise timekeeper, they derived analytically that the inter-tap intervals follow an ARMA(1,1) process with an autoregressive coefficient $(1-\kappa)$ governed by the correction gain $\kappa$ and a negative lag-1 autocorrelation reflecting the differencing of the shared motor delay across consecutive intervals (see section 2.6 for more details). This framework provides a rigorous account of the short-range serial dependence observed under synchronization but, owing to its white-noise generator, cannot reproduce the long-memory fluctuations. Torre and Delignières [25] resolved this limitation by replacing the white-noise timekeeper with a fractal (1/f) generator, yielding a model that spans both regimes. In self-paced conditions, no metronome reference is defined and the fractal timekeeper alone shapes the inter-response intervals, which remain persistent ($d > 0$). In synchronization conditions, in contrast, some active correction is mechanically required to prevent the cumulative asynchrony from diverging. This bounding constraint on the asynchrony series forces the inter-response intervals to be anti-persistent at the output ($d < 0$). The empirical diagnostic of an active correction loop is therefore a negative fractional differencing parameter on the observed series. Delignières and Torre [9] subsequently confirmed the same mechanism in the original Hausdorff [4] gait data under metronomic walking.

We extend this architecture in two directions. First, we argue that it operates not only under rhythmic cueing of stride time but also under the non-rhythmic positional constraint imposed by a constant treadmill belt speed. Second, we place the claim on a formal statistical footing: rather than selecting a BIC-best model, we apply the full eight-model ARFIMA family by Schwarz weights and estimate every parameter under Bayesian model averaging, yielding a direct test of whether the observed correlation structure requires a fractional differencing component at all.

Five more specific contributions follow from this framework. (i) We establish that cued gait anti-persistence is a genuine fractional phenomenon rather than a short-memory ARMA artifact, by showing that long-memory models prevail under Bayesian model averaging in both persistent and anti-persistent conditions across all three datasets. (ii) We quantify a systematic DFA overestimation bias of +0.25 to +0.34 α units attributable to short-memory components that DFA conflates with d and potentially to a bias related to finite-time limitation of the ML-ARFIMA approach. (iii) We interpret the ARFIMA parameters within the sensorimotor synchronization framework [25]: the fractional differencing parameter d as the primary signature of correction activity (d < 0 indicating an active loop bounding the cumulative asynchrony or positional drift), the autoregressive coefficient φ as a qualitative indicator of residual short-memory structure, and the moving-average coefficient θ as a marker of the motor delay [28]. (iv) We extend this mapping from rhythmic cueing of stride time to stride-speed regulation on a fixed-speed treadmill, where the positional constraint acts as a non-rhythmic analogue of the metronome reference. (v) We characterize empirically how series length constrains ARFIMA inference in gait data: long-memory family detection is robust even at T = 256 strides for anti-persistent conditions, while reliable discrimination of model structure requires progressively longer recordings.

## 2. Theoretical framework

This section presents the mathematical foundations underlying our analysis. We first define long-memory processes and the ARFIMA model family (Sections 2.1–2.2), then describe the model selection and parameter estimation strategy (Section 2.3) based on Bayesian model averaging (Section 2.4). Section 2.5 introduces detrended fluctuation analysis as a cross-validation tool. Finally, Section 2.6 connects the ARFIMA parameters to the sensorimotor synchronization architecture of gait control, generating testable predictions for the empirical analysis.

### 2.1. Long-memory processes and fractional differencing

A covariance-stationary process $\{X_t\}$ shows long-range dependence (long memory) if its autocovariance function decays as a power law rather than exponentially. Specifically, the autocovariances satisfy:

$$\gamma(k) \sim C \cdot |k|^{2d-1} \quad \text{as } k \to \infty, \quad d \neq 0 \tag{1}$$

where $d$ is the fractional differencing parameter and $C$ is a slowly varying constant [19,20], and $\gamma(k) = Cov(X_t, X_{t+k})$ denotes the autocovariance at lag $k$. When $d > 0$, the autocovariances are positive and their sum diverges, that is, the process is *persistent*, meaning that an increment in one direction tends to be followed by further increments in the same direction. When $d < 0$, the autocovariances at all lags $k \geq 1$ are uniformly negative and their sum converges to zero, the process is *anti-persistent*, meaning that increments in one direction tend to be followed by reversals. This power-law decay contrasts sharply with the exponential decay of autocovariances in short-memory (ARMA) processes, where temporal correlations vanish after a few lags [29].

The mechanism that generates long-memory structure is the fractional differencing operator $(1-B)^d$, where $B$ is the backshift operator ($BX_t = X_{t-1}$). This operator generalizes integer differencing to fractional orders:

$$(1-B)^d = \sum_{k=0}^{\infty} \pi_k \, B^k, \qquad \pi_k = \frac{\Gamma(k-d)}{\Gamma(k+1)\,\Gamma(-d)} \tag{2}$$

where $\Gamma(\cdot)$ denotes the gamma function [20]. For $d \in (-0.5,\, 0.5)$, the resulting process is stationary and invertible. The weights $\pi_k$ decay hyperbolically as $k \to \infty$, which is the source of the long-memory property: every past innovation contributes to the current value with a weight that diminishes slowly but never reaches zero.

The general ARFIMA$(p, d, q)$ model combines fractional differencing with short-memory autoregressive (AR) and moving-average (MA) polynomials:

$$\varphi(B)\,(1-B)^d\,X_t = \theta(B)\,\varepsilon_t \tag{3}$$

where $\varphi(B) = 1 - \varphi_1 B - \cdots - \varphi_p B^p$ is the AR polynomial, $\theta(B) = 1 + \theta_1 B + \cdots + \theta_q B^q$ is the MA polynomial, and $\varepsilon_t \sim \mathrm{WN}(0, \sigma^2)$ [19,20]. This model nests both long-memory ($d \neq 0$) and short-memory ($d = 0$) processes. The AR and MA parameters capture exponentially decaying short-range dynamics, while $d$ captures the hyperbolically decaying long-range structure. These two components are estimated simultaneously, allowing the model to disentangle long-memory from short-memory contributions. Recovering the parameters ($d$, $\varphi$, $\theta$, $\sigma^2$) jointly from a finite observation requires an estimator that exploits the full covariance structure of the ARFIMA process; the exact maximum likelihood procedure used here is derived in Section 2.3.

In the spectral domain, the ARFIMA process has a spectral density that behaves near frequency zero as:

$$f(\lambda) \sim C_f \cdot |\lambda|^{-2d} \quad \text{as } \lambda \to 0 \tag{4}$$

The spectral density at zero frequency measures how much variance the process carries at arbitrarily long timescales. Persistent processes ($d > 0$) show a spectral pole at zero frequency ($1/f$ noise), reflecting the concentration of variance at long timescales. Anti-persistent processes ($d < 0$) exhibit a spectral zero, indicating that long-timescale variance is suppressed relative to short-timescale fluctuations [29].

The relationship between spectral density and autocovariance structure can be made precise. For any stationary process, the spectral density at zero frequency equals the sum of all autocovariances [20,29]

$$f(0) = \frac{1}{2\pi} \sum_{k=-\infty}^{\infty} \gamma(k) \quad (5)$$

Combining this identity with the spectral behavior of FI processes (Eq. 4) yields three regimes: when $d > 0$, $f(0)$ diverges and the autocovariance sum is infinite (persistence); when $d = 0$, $f(0)$ is finite and the autocovariances vanish at all nonzero lags (white noise); when $d < 0$, $f(0) = 0$ and the autocovariance sum must equal zero exactly. In the anti-persistent regime, this means that the positive process variance $\gamma(0)$ must be precisely balanced by the cumulative negative autocovariance at all lags $k \geq 1$, which are uniformly negative and decay in absolute value hyperbolically (Eq. 1 and Fig. 1F). This cancellation is both necessary and sufficient for $d < 0$ in the FI family.

This asymmetry between persistent and anti-persistent long memory has a structural origin. Granger [30] showed that aggregating (superposing) many independent short-memory AR(1) processes whose autoregressive coefficients are drawn from a sufficiently spread distribution produces persistent long memory ($d > 0$) in the aggregate, because the mixture of exponentially decaying autocorrelation functions at different rates yields a net hyperbolic decay. This result is significant because it means that any system composed of interacting subprocesses operating at heterogeneous timescales (a description that fits hierarchical neural control networks [1,31]) can generate $1/f$ persistence without requiring any individual component to carry long-range temporal structure. The aggregation mechanism, however, produces exclusively persistent long memory; it cannot generate the exact autocovariance cancellation required by Eq. (5) [29,30]. Anti-persistent long memory therefore requires a qualitatively different generative process. In the sensorimotor context developed in Section 2.6, this process is the active error-correction loop that transforms the persistent output of a fractal timekeeper into anti-persistent inter-response intervals [25]. The rarity of anti-persistence across natural systems follows directly from the stringency of Eq. (5): persistence needs only a distribution of timescales; anti-persistence demands a mechanism that enforces the spectral zero.

This spectral characterization links the ARFIMA framework to the broader 1/f noise literature in physiology [31] and physics [32]. The parameter $d$ is the single quantity that separates long-memory from short-memory dynamics, and its sign determines whether the process is persistent, anti-persistent, or uncorrelated. The central inferential question for any empirical time series is whether $d$ differs significantly from zero, that is, whether the observed autocorrelation structure requires a long-memory component at all, or whether a short-memory ARMA model suffices. The model selection framework we use to answer this question is described next.

## 2.2. The ARFIMA model family

Fig. 1 illustrates the contrast between persistent and anti-persistent dynamics using simulated fractional Gaussian noise series (N = 1200). In the persistent case ($d = +0.30$, left column), the raw signal shows slowly wandering fluctuations (Panel A), the integrated profile drifts extensively (Panel C), the autocorrelation function decays as a power law that remains positive over 100 lags (Panel E). In the anti-persistent case ($d = -0.30$, right column), the signal appears rougher and more rapidly oscillating (Panel B), the integrated profile stays confined near zero (Panel D), the autocorrelation function is uniformly negative with hyperbolic decay and is concentrated within the first few lags (Panel F). In both cases, the power-law decay of the autocorrelation function (orange curves in Panels E–F) diverges from the exponential decay of a matched AR(1) process (dashed green curves), confirming that neither condition is adequately described by a short-memory model. Empirically, unconstrained walking produces $d \approx +0.17$ and metronomic walking produces $d \approx -0.48$ [4,9].

To test whether gait fluctuations require long-memory components, we define a family of eight nested models spanning the full ARFIMA(1,$d$,1) model space. Fig. 2 (Panel B) presents these models and their hierarchical structure. The four short-memory specifications (WN, AR, MA,

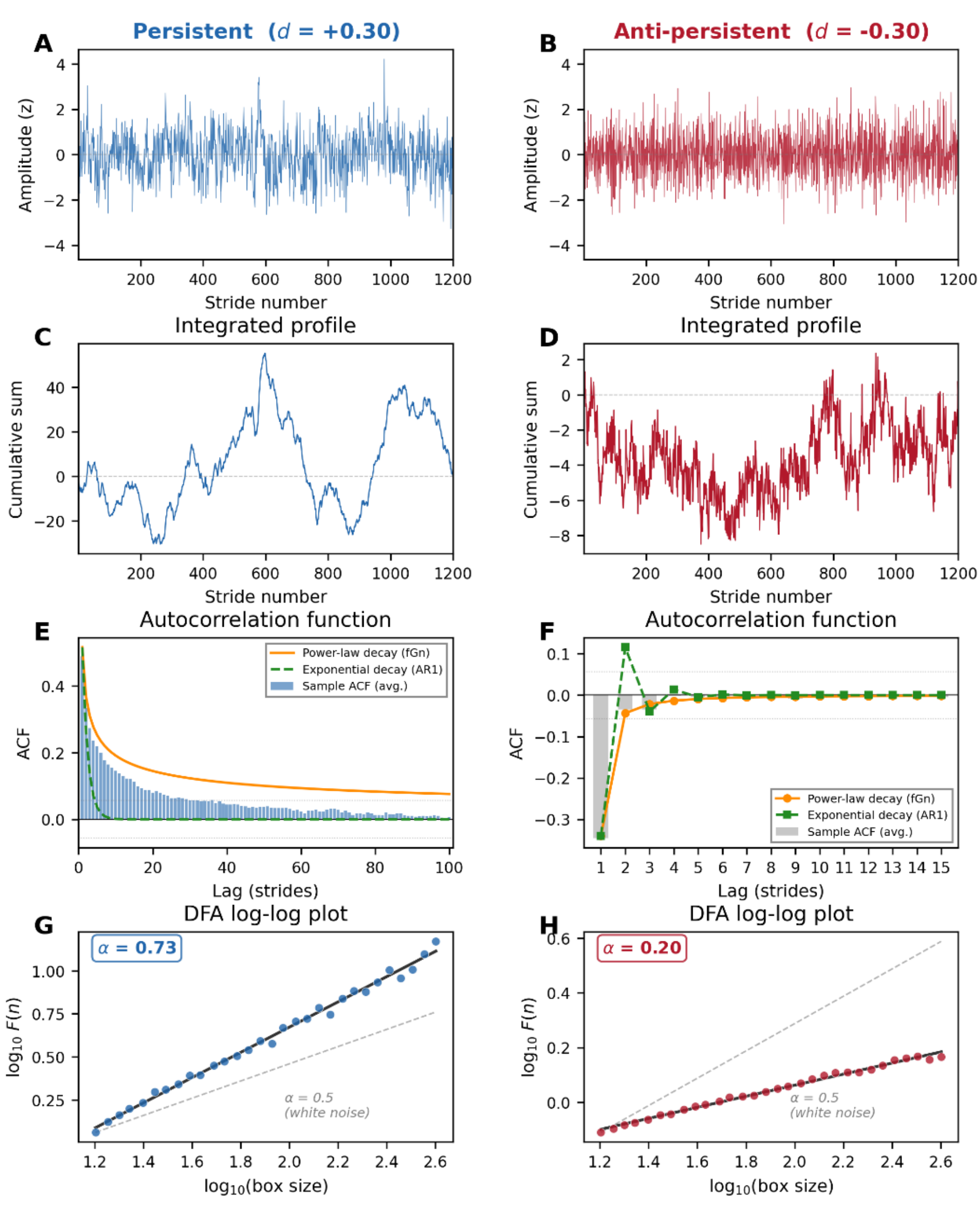

***Fig. 1. Persistent and anti-persistent long-range correlations in simulated stride time series.***

*Fractional Gaussian noise (fGn) was simulated with* $d = +0.3$ *(persistent, left column, blue) and* $d = -0.3$ *(anti-persistent, right column, red);* $N = 1200$ *strides.* ***(A, B)*** *Raw time series (z-scored).* ***(C, D)*** *Integrated profiles (cumulative sum of mean-centered series).* ***(E)*** *Autocorrelation function (ACF) of the persistent process, lags 1–100. Bars show the ensemble-averaged sample ACF over 30 independent realizations; solid orange line: theoretical fGn power-law decay (*$\gamma k^{2d-1}$*); dashed green line: ACF of a matched AR(1) process with identical lag-1 autocorrelation.* ***(F)*** *ACF of the anti-persistent process, lags 1–15. Both theoretical models share the same lag-1 autocorrelation (*$-0.34$*) but diverge sharply from lag 2 onward.* ***(G, H)*** *Detrended fluctuation analysis (DFA-1) log–log plots. Solid black line: fitted scaling exponent* $\alpha$*; dashed gray line:* $\alpha$ *reference (white noise). Persistence yields* $\alpha$ *(above 0.5), anti-persistence yields* $\alpha$ *(below 0.5).*

ARMA) set $d = 0$ and test whether the temporal correlations can be fully accounted for by exponentially decaying processes. The four long-memory specifications (FI, ARFI, FIMA, ARFIMA) include $d$ as a free parameter. The central inferential comparison is between these two groups: do the data require a fractional differencing component, or can the observed correlation structure be explained by short-memory dynamics alone?

We restrict the model space to $p, q \in \{0,1\}$ for three reasons. First, stride interval series of $T = 256$–$1200$ do not support reliable estimation of higher-order ARMA parameters [33,34]. Second, the ARFIMA($1,d,1$) model space corresponds to the standard framework introduced by Wagenmakers et al. [35] and applied to movement sciences by Torre et al. [17]. Third, $p = q = 1$ captures the dominant short-memory dynamics predicted by sensorimotor synchronization theory (Section 2.5).

## 2.3. Parameter estimation by exact maximum likelihood

Estimating the parameters (d, φ, θ, σ²) of each ARFIMA model jointly, and thereby disentangling the long-memory component d from the short-memory components (φ, θ), requires an estimator that exploits the full covariance structure of the process. We use the exact maximum likelihood (ML) procedure of Sowell [36]: each of the eight models in our framework is fitted to the observed series by this method.

Under the assumptions of stationarity ($|d| < 1/2$), invertibility, AR/MA polynomial roots strictly outside the unit circle, and Gaussian innovations $\varepsilon_t \sim iid\, \mathcal{N}(0, \sigma^2)$, the observed sample $Z_T = (z_1, z_2, \ldots, z_T)^\top$ is multivariate normal with zero mean and covariance matrix Σ:

$$Z_T \sim \mathcal{N}_T(0, \Sigma), \qquad p(Z_T \mid \beta) = (2\pi)^{-T/2}\, |\Sigma(\beta)|^{-1/2}\, exp(-1/2\; Z_T^\top\, \Sigma(\beta)^{-1}\, Z_T) \tag{6}$$

where $\beta = (d, \varphi, \theta, \sigma^2)$ collects the model parameters. Stationarity implies that the covariance matrix is Toeplitz: $\Sigma_{ij} = \gamma(|i-j|)$, where $\gamma(\cdot)$ is the autocovariance function of the process.

The central contribution of Sowell [36] is a closed-form parameterization of $\gamma(s)$ as a function of $(d, \varphi, \theta, \sigma^2)$, obtained without infinite-series truncation. Starting from the spectral density of the ARFIMA process,

$$f_z(\lambda) = \frac{\sigma^2}{2\pi} \frac{\left|\theta\left(e^{i\lambda}\right)\right|^2}{|\varphi(e^{i\lambda})|^2} \left|1 - e^{i\lambda}\right|^{-2d}, \qquad \lambda \in [-\pi, \pi] \tag{7}$$

and inverting through

$$\gamma(s) = \int_{-\pi}^{\pi} f_z\,(\lambda)\, e^{i\lambda s}\, d\lambda \tag{8}$$

Sowell derives an exact expression for $\gamma(s)$ in terms of Gauss hypergeometric functions of the autoregressive roots and the differencing parameter $d$. This representation makes the entire Toeplitz covariance matrix $\Sigma(\beta)$ computable to machine precision for any admissible $\beta$.

The exact log-likelihood follows directly:

$$lnL(\beta) = -T/2\ ln(2\pi) - 1/2\ ln|\Sigma(\beta)| - 1/2\ Z_T^{\top}\, \Sigma(\beta)^{-1}\, Z_T \tag{9}$$

and is evaluated at computational cost $O(T^2)$ via Levinson recursion applied to the Toeplitz Cholesky factorization [36]. Maximum-likelihood estimates $\hat{\beta}_{ML}$ are obtained by numerical optimization of Eq. (9). The same maximized likelihood feeds directly into the BIC-based model selection described in Section 2.4, ensuring internal consistency between estimation and selection.

Under the regularity conditions stated above, Dahlhaus [37] established that $\hat{\beta}_{ML}$ is consistent, asymptotically normal, and asymptotically efficient:

$$\sqrt{T}\left(\hat{\beta}_{ML} - \beta\right) \ \overset{d}{\rightarrow}\ \mathcal{N}\left(0, \mathcal{I}^{-1}(\beta)\right), \qquad T \rightarrow \infty \tag{10}$$

attaining the Cramér–Rao lower bound. In the long-series limit, exact ML therefore provides the most informative parametric estimator available for the ARFIMA family.

At finite series lengths, however, the small-sample behavior of $\hat{d}_{\mathrm{ML}}$ may depart from its asymptotic distribution. Monte-Carlo evidence indicates a potential negative bias on $\hat{d}$ when the series mean is unknown and is plug-in-estimated, with the bias amplified by coexisting short-memory ARMA components [38]. The magnitude of this finite-$T$ bias on the $(d, T,$ARMA$)$ configurations typical of gait recordings remains an open question to be investigated.

## 2.4. Model selection and Bayesian model averaging

Rather than selecting a single "best" model, we use Bayesian model averaging (BMA) to propagate model uncertainty into the parameter estimates [39]. The procedure consists of three steps: computing information criteria, converting them to model weights, and averaging across models.

For each model $M_k$ with $k$ free parameters fitted to $n$ observations, the Bayesian Information Criterion (BIC) is:

$$\mathrm{BIC}_k = -2\ln\left(\hat{L}_k\right) + k\ln(n) \tag{11}$$

where $\hat{L}_k$ is the maximized likelihood [40]. BIC penalizes model complexity more heavily than the Akaike Information Criterion (AIC), favoring parsimonious models. This conservatism is deliberate: if a long-memory model prevails despite the stronger BIC penalty, the evidence for long memory is robust.

BIC values are converted to approximate posterior model probabilities (Schwarz weights) via:

$$\Delta_k = \mathrm{BIC}_k - \mathrm{BIC}_{\min} \tag{12}$$

$$w_k = \frac{\exp(-\Delta_k/2)}{\sum_{j=1}^{8}\exp\left(-\Delta_j/2\right)} \tag{13}$$

where $\mathrm{BIC}_{\min}$ is the smallest BIC across all eight models [35,41]. Each weight $w_k \in [0,1]$ and the weights sum to unity. They approximate the posterior probability $P(M_k \mid \text{data})$ under equal prior model probabilities [42].

We define the long-memory weight as the aggregate probability assigned to models containing a fractional differencing parameter:

$$W_{\text{long}} = w_{\text{FI}} + w_{\text{ARFI}} + w_{\text{FIMA}} + w_{\text{ARFIMA}} \tag{14}$$

$W_{\text{long}} > 0.5$ indicates that the data favour long memory; $W_{\text{long}}$ near 1 indicates unambiguous long-memory classification.

The fractional differencing parameter is estimated via BMA:

$$d_{\text{BMA}} = \sum_{k=1}^{8} w_k \cdot \hat{d}_k \tag{15}$$

where $\hat{d}_k$ is the maximum likelihood estimate of $d$ under model $M_k$, with $\hat{d}_k = 0$ for short-memory models (WN, AR, MA, ARMA). This estimator is a shrinkage estimator: if short-memory models carry substantial weight, $d_{\text{BMA}}$ is pulled toward zero; if long-memory models dominate, $d_{\text{BMA}}$ reflects the estimated long-memory strength. The sign of $d_{\text{BMA}}$ determines whether the process is persistent ($d_{\text{BMA}} > 0$) or anti-persistent ($d_{\text{BMA}} < 0$), and its magnitude quantifies the strength of the long-range correlation structure [35].

The same averaging logic extends to the short-memory parameters. We define BMA-weighted autoregressive and moving-average coefficients as:

$$\varphi_{\text{BMA}} = \sum_{k=1}^{8} w_k \cdot \hat{\varphi}_k \tag{16}$$

$$\theta_{\text{BMA}} = \sum_{k=1}^{8} w_k \cdot \hat{\theta}_k \tag{17}$$

where $\hat{\varphi}_k$ and $\hat{\theta}_k$ are the maximum likelihood estimates of the autoregressive and moving-average coefficients under model $M_k$, with $\hat{\varphi}_k = 0$ for models without an AR component (WN, MA, FI, FIMA) and $\hat{\theta}_k = 0$ for models without an MA component (WN, AR, FI, ARFI). Like $d_{\text{BMA}}$, these are shrinkage estimators: $\varphi_{\text{BMA}}$ and $\theta_{\text{BMA}}$ are pulled toward zero when models lacking the corresponding parameter carry substantial Schwarz weight.

## 2.5. Detrended fluctuation analysis

Detrended fluctuation analysis (DFA) provides a model-free estimate of the scaling behavior of a time series, serving as an independent cross-validation of the ARFIMA results. Originally developed for non-stationary physiological signals [2,3], DFA has become the standard method for quantifying fractal dynamics in gait research.

Given a time series $\{X_t\}$ of length $N$, the algorithm proceeds in four steps. First, the series is integrated by computing the cumulative sum of mean-centred values:

$$Y(k) = \sum_{t=1}^{k} (X_t - \bar{X}), \qquad k = 1, \dots, N \tag{18}$$

Second, the integrated profile $Y(k)$ is divided into non-overlapping segments of size $n$, and a polynomial of order $m$ is fitted within each segment to remove local trends (DFA-$m$). Third, the root-mean-square fluctuation function is computed:

$$F(n) = \sqrt{\frac{1}{N} \sum_{k=1}^{N} [Y(k) - Y_n(k)]^2} \tag{19}$$

where $Y_n(k)$ is the local polynomial fit. Fourth, the scaling exponent $\alpha$ is estimated as the slope of the linear regression of $\log F(n)$ versus $\log(n)$:

$$F(n) \sim n^{\alpha} \tag{20}$$

The scaling exponent $\alpha$ classifies the temporal correlation structure: $\alpha = 0.5$ indicates uncorrelated noise (white noise); $\alpha > 0.5$ indicates persistent long-range correlations; $\alpha < 0.5$ indicates anti-persistent long-range correlations; $\alpha = 1.0$ corresponds to $1/f$ noise; and $\alpha = 1.5$ corresponds to Brownian motion [2,3].

Panels G and H of Fig. 1 illustrate the DFA log-log plot for a persistent ($d = +0.30, \alpha = 0.73$) and an anti-persistent ($d = -0.30, \alpha = 0.20$) fractional noise, respectively. In both cases, the scaling relationship is clearly linear over the full range of box sizes, and the slopes diverge markedly from the $\alpha = 0.5$ reference line (white noise), confirming the presence of long-range correlations.

For a pure fractionally integrated (FI) process, the DFA scaling exponent is related to the fractional differencing parameter by:

$$\alpha = d + 0.5 \tag{21}$$

This relationship provides a direct bridge between the ARFIMA-derived $d$ and the DFA scaling exponent [2,43]. However, Eq. (21) holds only for pure fractional noise. When short-memory AR or MA components are present (as in ARFI, FIMA, or full ARFIMA models) DFA $\alpha$ is influenced by both the long-memory component $d$ and the short-memory parameters $\varphi$ and $\theta$. This creates a systematic discrepancy: DFA $\alpha$ reflects the *combined* effect of all temporal correlations, whereas $d_{\mathrm{BMA}}$ isolates the long-memory component. In practice, ARMA components tend to inflate DFA $\alpha$ relative to $d + 0.5$, producing a systematic positive bias [17,44,45]. This is one of the central arguments for using ARFIMA rather than DFA alone: ARFIMA decomposes what DFA conflates.

## 2.6. Sensorimotor synchronization and ARFIMA parameter mapping

We now map the ARFIMA parameters (d, φ, θ) onto the sensorimotor control architecture of gait, using the tapping-synchronization model of Torre and Delignières [25] as the reference framework. The mapping gives each estimated parameter a mechanistic role: d carries the principal signature of correction activity: its sign separates the self-paced regime (d > 0, fractal timekeeper alone) from the synchronization regime (d < 0, active correction enforcing a spectral zero). In addition, φ and θ are qualitative markers of the residual short-memory structure that the correction process and motor delay impart on the observed series. The remainder of this section develops the mapping step by step.

The classical Wing–Kristofferson model [46] decomposes the inter-response interval $I_n$ into a central timekeeper interval $C_n$ and motor implementation delays $M_n$ :

$$I_{\mathrm{n}} = C_{\mathrm{n}} + M_{n+1} - M_{\mathrm{n}} \tag{22}$$

In the original formulation, both $C_n$ and $M_n$ are independent white-noise processes. This produces a characteristic autocorrelation structure in the observed intervals: $\gamma(1) = -Var(M)$ from the shared motor delay at the boundary between consecutive intervals, and $\gamma(k) = 0$ for all $k \geq 2$. This is an MA(1) process.

Vorberg and Schulze [28,47] embedded the Wing–Kristofferson architecture into a synchronization paradigm by adding a linear phase-correction loop. Assuming a white-noise timekeeper ($C_n$ i.i.d. with $\mathrm{E}[C_n] = \mu_C$ and $\mathrm{Var}(C_n) = \sigma_C^2$) and white-noise motor delays ($M_n$ i.i.d. with $\mathrm{Var}(M_n) = \sigma_M^2$), they postulated that the subject corrects each synchronization error $A_n$ by subtracting a fraction $\kappa$ of it from the next produced interval. Under this assumption, the asynchrony (the time difference between the produced event and metronome beat) sequence obeys the first-order stochastic difference equation [28, Eq. 7]:

$$A_{n+1} = (1-\kappa)\,A_n + H_n \tag{23}$$

where $H_n = C_n + M_{n+1} - M_n - \tau$ collects the timekeeper, motor, and metronome noise contributions, and $\tau = E[C_n]$ is the metronome period. The asynchronies therefore form a first-order autoregressive process with coefficient $(1-\kappa)$, asymptotically stationary for $0 < \kappa < 2$ [28, Theorem 3.2]. Combined with the kinematic identity $I_n = \tau + A_{n+1} - A_n$, this dynamics implies that the inter-response intervals $I$ follow an ARMA(1,1) process with autoregressive coefficient $(1-\kappa)$ and a moving-average coefficient whose magnitude and sign depend on the ratio of motor to timekeeper variance [28, Theorem 3.5]. Although this framework rigorously accounts for the short-range serial dependence of synchronized intervals (including the negative lag-1 autocorrelation produced by the differenced motor delay) it cannot reproduce the long-memory fluctuations observed in either self-paced or metronome-paced sensorimotor production, because the assumed white-noise timekeeper carries no temporal memory beyond the next interval.

Torre and Delignières [25] proposed a critical amendment: replacing the white-noise timekeeper with a fractal timekeeper whose intervals $C_n$ show genuine long-range dependence, characterized by a fractional differencing parameter $d_t > 0$, where the subscript distinguishes the timekeeper's intrinsic parameter from the observed parameter d estimated from the output series. The motor delays remain white noise. Under self-paced conditions, $I_n$ (the inter-response interval) is the sum of a fractal process and an MA(1) component from the shared motor delay (Eq. 22), yielding a theoretical FIMA(0,d,1) structure. The phase-correction loop (Eq. 23) continues to operate, with $C_n$ now a fractal process; combining it with the Wing–Kristofferson decomposition (Eq. 22) yields the observed inter-response interval as:

$$I_n \;=\; C_n \;+\; (M_{n+1} - \; M_n) \;-\; \kappa \;\cdot\; A_n \tag{24}$$

Eq. (25) reveals the mechanism of the persistent-to-anti-persistent transition. With no correction, $I_n$ reduces to the Wing–Kristofferson model (Eq. 22) and inherits the persistent fractal structure of $C_n$. As $\kappa$ increases toward 1, the correction term $-\kappa \cdot A_n$ reverses the temporal correlations: the timekeeper still generates persistent fluctuations, but the correction loop overcompensates at each stride, transforming the output from persistent to anti-persistent [25]. This reversal is not incidental but necessary. The asynchrony $A_n$ is the running sum of the deviations $(I_n - \tau)$; for $A_n$ to remain bounded (which synchronization requires) the spectral density of $I_n$ must vanish at zero frequency (Eq. 5), forcing d < 0 in the output series. A persistent output (d > 0) would generate unbounded asynchrony drift, making synchronization physically impossible. Delignières and Torre [9] confirmed that the same mechanism applies to the Hausdorff et al. [4] gait data: metronomic walking produced anti-persistent stride intervals ($d \approx -0.33$ to $-0.44$) and persistent asynchrony series, exactly as predicted.

In summary, the Torre model [25] maps directly onto the ARFIMA parameter structure. The fractional differencing parameter d (Eq. 3) corresponds to the net long-range structure after correction: positive d when the fractal timekeeper dominates, negative d when the correction overwhelms the timekeeper's intrinsic persistence. The autoregressive coefficient $\varphi$ is a qualitative indicator of incomplete asynchrony correction. Under the bare Vorberg–Schulze model (Eq. 23), the inter-response intervals follow an ARMA(1,1) process whose AR(1) coefficient equals $(1 - \kappa)$: $\varphi \to 0$ corresponds to asymptotically perfect correction (each asynchrony fully neutralized before the next interval) and $\varphi$ close to 1 corresponds to weak correction. Under the Torre fractal extension (Eqs. 24–25), Torre and Delignières [25] showed through simulation calibration that the long-memory component of the timekeeper biases the parametric estimate of $(1 - \kappa)$ downward, so that $\varphi$ is best read as a qualitative marker of an active correction loop rather than a quantitative estimate of the correction gain. The moving-average coefficient $\theta$ reflects two superposed contributions: the Wing–Kristofferson motor delay at the boundary between consecutive intervals (Eq. 22), which by itself produces a negative $\theta$ at the inter-response level, and, under synchronization, a residual signature of the correction process whose magnitude depends on the ratio of motor to timekeeper variance and on the correction strength $\kappa$ [28, Theorem 3.5].

## 3. Datasets and analysis pipeline

We applied the eight-model ARFIMA family and BMA estimation procedure described in Section 2 to stride interval time series from three independent datasets. The datasets span different walking environments (overground and treadmill), cueing modalities (auditory, visual, isochronous pacing), and spatial constraint levels, providing a broad test of the ARFIMA approach. Table 1 summarizes their key characteristics.

**Table 1.** Characteristics of the datasets.

| | Hausdorff et al. (1996) | Terrier (2016) | Roerdink et al. (2019) |
|---|---|---|---|
| ***Study characteristics*** | | | |
| Participants | 10 | 36 | 24 |
| Sex (M/F) | 10/0 | 14/22 | 5/19 |
| Age (yr) | 21.7 (18–29) | 33 ± 10 | 23 ± 2 |
| ***Protocol*** | | | |
| Walking environment | Overground | Treadmill | Treadmill |
| Equipment | Force-sensitive insoles | C-Mill | C-Mill |
| Walking speed | 3 self-selected speeds | 4.0 ± 0.6 km/h | 4.3 ± 0.4 km/h |
| Design | 2 cueing × 3 speeds | 3 cueing conditions | 3 walking areas × 2 cueing |
| Constraint type | Temporal:<br>isochronous metronome | Temporal:<br>isochronous metronome<br>Spatial:<br>visual stepping stones<br>Positional:<br>rope | Temporal:<br>isochronous metronome<br>Positional:<br>walking area boundaries |
| Conditions (total) | 6<br>(3 speeds × with/without metronome) | 3<br>(NC, AC, VC) | 6<br>(3 walking areas × with/without metronome) |
| Duration per condition | ~1 h | ~10 min | ~5 min |
| ***Time series*** | | | |
| Recorded length | ~1200–3500 strides | 500 strides | 280 strides |
| Analysis length | Tier 1: L = 1200<br>Tier 2: L = 600<br>(non-overlapping bouts) | T = 500 | T = 256 |
| Gait parameters [a] | **ST** | **ST**, SL, **SS** | ST, SL, **SS** |
| Series analyzed | 60 (Tier 1, 10 × 3 cond.)<br>206 bouts (Tier 2, 10 × 3 cond. × variable) | 216<br>(36 × 3 cond. × 2 param.) | 144<br>(24 × 3 positional × 2 temporal cond.) |
| Data source | PhysioNet [b] | Supplementary Material S1 [c] | Roerdink et al. (2019) [d] |

Values are mean ± SD unless stated otherwise. NC = no cueing; AC = auditory cueing; VC = visual cueing.

[a] Bold = parameters included in analysis. ST = stride time; SL = stride length; SS = stride speed.

[b] Stride interval time series from the PhysioNet Long-Term Gait Dynamics database [48], record "umwdb" v1.0.0.

[c] Raw center-of-pressure trajectories available on Figshare [49]. Stride time series are provided in the Zenodo reproduction archive.

[d] Stride time series provided as supplementary .mat files in Roerdink et al. [14].

## 3.1. Hausdorff (1996): overground walking

The first dataset comes from the PhysioNet "Gait Dynamics in Healthy Subjects" database [48] and contains overground walking data originally reported by Hausdorff et al. [4]. Ten healthy young adults (age 18–29 years) each completed six continuous walking sessions: three unconstrained (at self-selected slow, normal, and fast speeds) and three metronomic (synchronizing heel strikes to an isochronous metronome set to the corresponding preferred cadence). Stride intervals were recorded via force-sensitive insoles. Each unconstrained session lasted approximately one hour, yielding 2040–3822 strides. Metronomic sessions were shorter (0.5 h, 1210–1956 strides) because the metronome frequency was derived from the preceding unconstrained walks and the shorter duration was deemed sufficient for correlation analysis.

The raw series lengths vary by a factor of three across conditions and subjects, which creates a confound for both ARFIMA estimation reliability and DFA scaling behavior [33]. We addressed this by segmenting the raw series into non-overlapping bouts of fixed length. Tier 1 truncates every series to exactly 1200 strides (the longest length achievable in all 60 subject×condition cells), yielding one bout per cell for a uniform cross-condition comparison. Tier 2 segments each series into bouts of 600 strides (discarding any remainder), producing 2–7 bouts per cell and a total of 206 bouts across all subjects and conditions. Tier 2 serves as the primary dataset for mixed-effects modeling; Tier 1 provides a secondary analysis at maximum uniform length.

A third tier (T3) extends the analysis to T = 2000 strides in the unconstrained conditions only, since the metronomic bouts are structurally too short to reach this length. T3 is constructed by truncating the unconstrained series of each subject at 2000 strides, yielding N = 10 subjects × 3 speeds = 30 series. T3 is designed specifically to test whether the short-memory parameters detected at T = 1200 (Section 4.1.2) are robust to further series-length extension, and whether the moving-average component predicted by the sensorimotor synchronization model (Section 2.6) receives statistical support from BIC at longer T. The complete T3 results are reported in Supplementary Table S1. Only stride time (ST) is available across all three tiers.

## 3.2. Terrier (2016): treadmill cueing

The second dataset was originally reported by Terrier [13] and consists of treadmill walking data from 36 healthy adults (21 women, 15 men; age 38 ± 13 years). Participants walked on an instrumented treadmill (C-Mill, ForceLink, Culemborg, The Netherlands) equipped with an embedded force platform (sampling at 500 Hz) spanning the full walking area. The platform measured vertical ground reaction force at 500 Hz. The belt speed was fixed at each participant's preferred walking speed.

Three cueing conditions were administered in randomized order: no cueing (NC), in which participants walked without external pacing while maintaining their position relative to a fixed rope landmark; auditory cueing (AC), in which an isochronous metronome matched to the participant's preferred cadence paced heel strikes; and visual cueing (VC), in which stepping stones were projected onto the treadmill belt at intervals corresponding to the participant's preferred stride length and synchronized with the belt speed. Stride time (ST) and stride length (SL) were extracted from the longitudinal center-of-pressure signal using a heel-strike detection algorithm validated for large treadmill-embedded force platforms [13,50]. stride speed (SS) was computed as the ratio SL/ST. In each condition, 500 consecutive gait cycles were recorded. We analyzed ST and SS only (216 series: 36 subjects × 3 cueing conditions × 2 parameters); SL is not used in either of the analysis tracks defined in Section 3.4.

## 3.3. Roerdink (2019): walking area constraint

The third dataset was originally reported by Roerdink et al. [14] and provides a graded manipulation of positional constraint tightness. Twenty-four healthy young adults (19 women; age 23 ± 2 years) walked on the same C-Mill treadmill model used in the Terrier study [13]. Walking areas of three sizes were projected onto the belt: small (120% of each participant's preferred stride length), intermediate (215%), and large (full 3-m belt). Each walking area was tested with and without isochronous acoustic pacing, yielding a 3 × 2 fully crossed design (six conditions total). Per condition, 280 strides were recorded and the first 20 discarded, leaving 256 strides for analysis. We analyzed SS only in six conditions, since SS is anti-persistent under the fixed-speed belt constraint regardless of pacing [10,12,13]. The stride-level time series were obtained from the supplementary materials of the original publication.

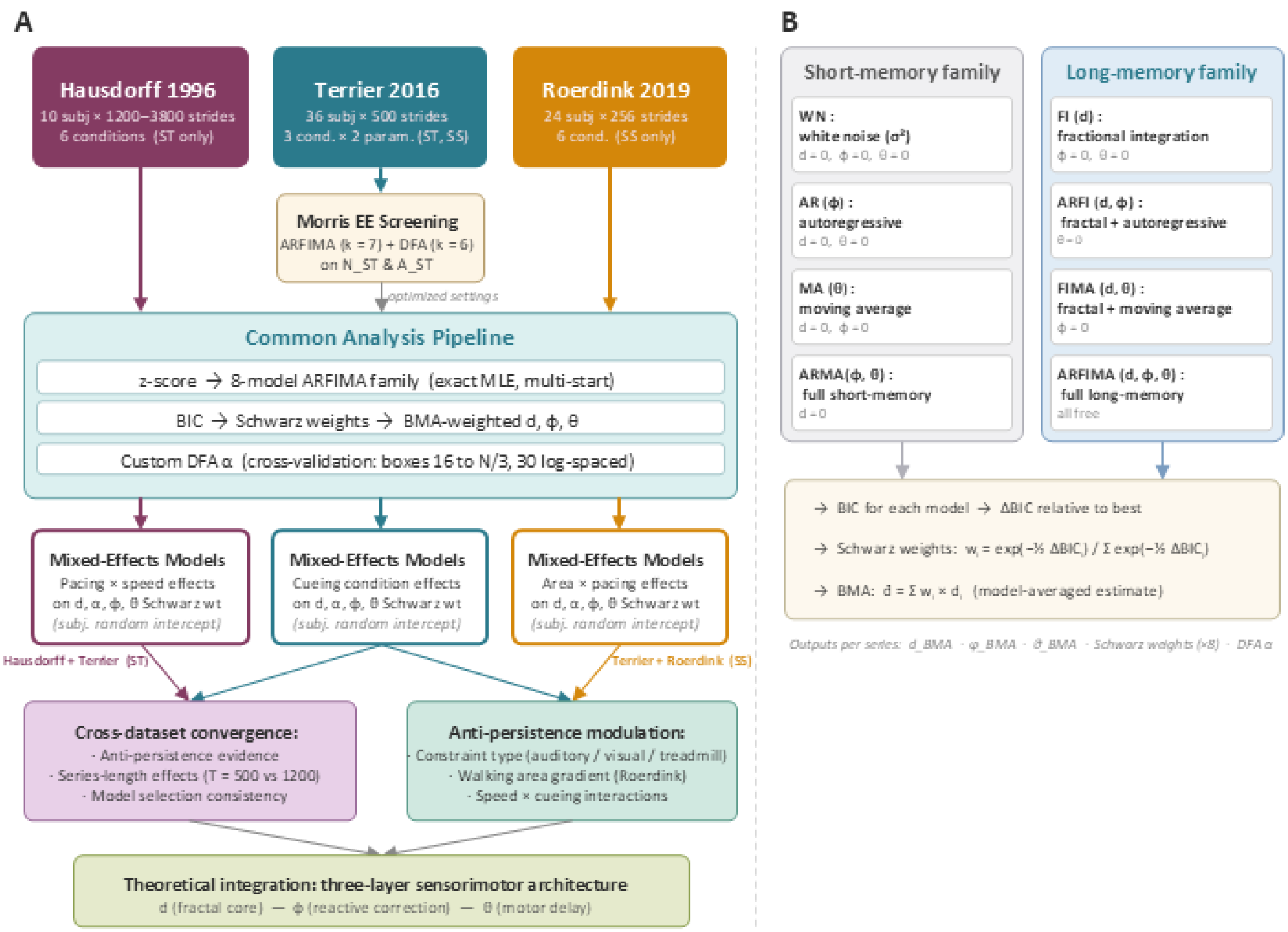

***Fig. 2. Analysis pipeline and ARFIMA model architecture.***

*(A) Overview of the multi-dataset analysis workflow. Schwarz weights: $w_i = exp(-½\ ΔBIC_i)/\sum_j exp(-½\ ΔBIC_j)$, where $ΔBIC_i$ is the difference between the BIC of model i and the best-fitting model.* ***(B)*** *The eight nested ARFIMA sub-models with their free parameters. d: fractional differencing parameter (long-range dependence); φ: autoregressive coefficient (residual short-memory structure); θ: moving-average coefficient (motor delay and residual correction); σ²: innovation variance. Abbreviations: ARFIMA, autoregressive fractionally integrated moving average; BIC, Bayesian information criterion; BMA, Bayesian model averaging; DFA, detrended fluctuation analysis; EE, Elementary Effects; MLE, maximum likelihood estimation; SS, stride speed; ST, stride time.*

### 3.4. Analysis organization

The research questions posed above require two distinct types of evidence: (1) that the persistent-to-anti-persistent shift in stride intervals under cueing reflects genuine long-memory dynamics rather than short-memory artifacts, and (2) that the strength of anti-persistence varies with the degree of external constraint. These two questions map onto different gait parameters and different dataset combinations, because not all stride parameters carry the same information in all experimental designs (Table 1, Fig.

2, Panel A). Stride time (ST) is the parameter for which long-range persistence is best established in unconstrained walking [4,9,13]; the Hausdorff and Terrier datasets both manipulate isochronous pacing within subjects and both record ST, making them the natural pair for testing the cueing-induced shift. Stride speed (SS), by contrast, is anti-persistent on any constant-speed treadmill regardless of cueing, because the fixed belt forces active speed regulation [10,13,15]; in the Terrier and Roerdink datasets, both collected on the same instrumented treadmill (C-Mill), SS anti-persistence is present in every condition, but its magnitude may vary with constraint type and tightness. This parameter is therefore the right probe for the second question: whether ARFIMA modeling resolves graded differences in anti-persistence that DFA $\alpha$ may miss. The remaining contributions (DFA and ML-ARFIMA finite-sample bias quantification, ARFI model dominance, and series length effects) are assessed jointly across both tracks, since they concern the behavior of the estimation pipeline itself rather than any single experimental manipulation. These considerations dictate a two-track analysis design.

The first track targets the cueing effect on stride time and draws on the Hausdorff and Terrier datasets. Both datasets share the same within-subject manipulation (unconstrained walking versus isochronous metronomic pacing) and both record the number of strides needed for reliable ARFIMA estimation (T = 500 and T = 600/1200, respectively). This track tests whether the persistent-to-anti-persistent shift in ST documented by DFA [4,8,13] survives formal model comparison. It also exploits the variable series length in the Hausdorff data (Tier 1: L = 1200; Tier 2: L = 600; T3 L = 2000) to assess how estimation reliability changes with recording duration.

The second track targets the modulation of anti-persistence strength and draws on the Roerdink and Terrier datasets, analyzing SS exclusively. Both datasets were collected on the same instrumented treadmill (C-Mill), eliminating equipment as a confound. SS is anti-persistent in all conditions because the constant belt speed forces active speed regulation [10,12,14,15]. The question is whether the *degree* of anti-persistence varies with constraint tightness. In the Roerdink data, the walking area manipulation provides three levels of spatial constraint (small, intermediate, large); in the Terrier data, the three cueing conditions (no cueing, auditory, visual) impose different temporal, spatial, and positional demands. This track tests whether ARFIMA $d$ captures graded changes that DFA $\alpha$ may not resolve, and whether the dominant ARFIMA model structure (the balance between fractal and short-memory components) shifts with constraint type.

Both tracks feed into the same estimation pipeline (Fig. 1): z-score normalization, eight-model ARFIMA fitting with BIC-based Schwarz weights and BMA-weighted $d$, and parallel DFA $\alpha$ computation as cross-validation. Within each dataset, condition effects are tested with linear mixed-effects models (subject as random intercept).

### 3.5. Parameter sensitivity screening

The analysis pipeline involves several tunable parameters, from preprocessing choices (detrending, outlier handling) to estimation settings (number of optimization restarts, mean estimation, parameterization family) and DFA configuration (polynomial order, box-size range, spacing). Before fixing the pipeline, we screened all of these using the Morris Elementary Effects method [51,52], a one-at-a-time global sensitivity design that classifies each parameter as influential, non-influential, or interactive at low computational cost. Two separate screenings were conducted on the Terrier stride time data (no cueing ST (persistent) and auditory cueing ST (anti-persistent), 36 series each): one for the ARFIMA estimation parameters ($k = 7$ factors) and one for the DFA parameters ($k = 6$ factors), with r = 20 trajectories drawn from 500 candidates and p = 6 levels per factor.

ARFIMA models were fitted with the arfima R package (v1.8-1), which implements exact maximum likelihood estimation for the full ARFIMA(p,d,q) family including all nested sub-models [33,53]. The package handles multi-modal likelihoods through multi-start optimization (controlled by the numeach parameter) and supports both the fractional differencing (fd) and fractional Gaussian noise (fGn) parameterizations (the lmodel parameter). Veenstra [33] documents the fitting engine in detail, including known boundary behavior when the fractional parameter d approaches -0.5 or +0.5, and the interaction between dynamic mean estimation (dmean) and mean-zero input series ([33] Chapter 6).

The screening identified lmodel as the single most influential factor for anti-persistent d estimates ($\mu* = 0.525$, more than double any other parameter): the fd parameterization ('d') produces anti-persistent estimates consistent with DFA cross-validation, whereas the fGn parameterization ('g') substantially attenuates them. This discrepancy arises because, although the fd and fGn spectral densities are asymptotically equivalent near zero frequency, they diverge at higher frequencies precisely where anti-persistent processes concentrate their distinctive spectral energy [29,33]. The number of optimization restarts (numeach) was the second key parameter, with a consistent positive effect on long-memory classification (+13.7 percentage points), indicating that fewer restarts miss global optima in multi-dimensional models. Enabling dynamic mean estimation (dmean = TRUE) on z-scored input caused boundary trapping in anti-persistent conditions ([33] Chapter 6). Preprocessing choices (detrending, outlier handling) were non-influential for both $d$ and $\alpha$. For DFA, polynomial order (DFA-1 vs. DFA-2) was the dominant parameter; box-size minimum and spacing showed moderate additive effects, while regression method was non-influential. Full screening results, including the Morris $\mu*$ and $\sigma$ plots for all parameters and outputs, are reported in Supplementary Material S2.

### 3.6. Estimation pipeline

Each stride series was z-score normalized (zero mean, unit variance) before model fitting. No polynomial detrending was applied; the Morris screening confirmed its negligible influence on $d$ and $\alpha$ in these data. The eight-model ARFIMA family (Eqs. (3), Section 2) was fitted to every series. Short-memory models used the standard ARMA parameterization (lmodel = 'n'); long-memory models used the fractional differencing parameterization (lmodel = 'd'). Based on the screening results, all fits used dmean = FALSE (fixed zero mean, appropriate for z-scored input). Multi-start optimization was set to numeach = c(5,5) for the ARFI, FIMA, and ARFIMA models, whose two- and three-dimensional parameter spaces are prone to multi-modal likelihoods [33]. The FI model used numeach = c(1,1) to circumvent a known package bug triggered at higher values. Short-memory models required no multi-start (c(1,0)). For each series, the pipeline returns the BIC for all eight models, the corresponding Schwarz weights (Eq. (13)), the BMA-weighted $d$ (Eq. (15)), and the individual model parameters ($\varphi$, $\theta$, $d$). BMA-weighted $\varphi$ and $\theta$ were computed following the same Burnham and Anderson [41] logic as $d_{BMA}$, with models lacking an AR or MA component contributing zero to the respective weighted average. Residual adequacy was assessed by applying a Ljung–Box test (lag = 10) to the residuals of the BIC-best model for each series [54].

DFA scaling exponents were computed in parallel as a cross-validation tool. We used a custom implementation following Almurad and Delignieres [55] with first-order polynomial detrending (DFA-1), box sizes from $n = 16$ to $n = N/3$, and 30 geometrically spaced points on the log–log plot.

The complete per-series ARFIMA and DFA output for all three datasets is archived in a dedicated Zenodo reproduction archive (https://doi.org/10.5281/zenodo.19676064). In addition, pipeline correctness was verified on 600 simulated ARFIMA series: the anti-persistent regime is recovered without bias, and a negative bias in short time series makes the empirical persistent d values reported here conservative lower bounds (see validation_report.md in the reproduction archive).

### 3.7. Statistical analysis

We analyzed condition effects on the ARFIMA and DFA outcomes using linear mixed-effects models (LMMs) and beta-family generalized linear mixed models (GLMMs), fitted separately for each dataset. Eight outcome variables were modeled per dataset. Continuous outcomes (the BMA-weighted fractional differencing parameter $d_{BMA}$, the DFA scaling exponent $\alpha$, and the BMA-weighted autoregressive and moving-average coefficients $\varphi_{BMA}$ and $\theta_{BMA}$) were fitted with LMMs estimated by restricted maximum likelihood (REML). BIC model-selection weights ($W_{long}$, $W_{ARFI}$, $W_{ARMA11}$), which are bounded on (0, 1), were fitted with beta GLMMs using a logit link; boundary values were squeezed to the open interval (0.001, 0.999) prior to fitting. We also computed the normalized Shannon entropy of the eight-model BIC weight vector, $H_{norm} = -\frac{\sum w_k \ln w_k}{\ln 8}$, as a model-selection certainty index (0 when a single model dominates; 1 for uniform weights) and included it among the continuous LMM outcomes.

All models included a subject random intercept. For the Hausdorff Tier 2 data (multiple bouts per subject), we also tested a random slope for the pacing factor on the primary outcomes ($d_{BMA}$, $\alpha$) via likelihood ratio test.

The fixed-effects structure was adapted to each dataset's experimental design. For the Hausdorff dataset, we specified a pacing × speed interaction with centered contrast codes (pacing: −0.5/+0.5; speed: linear), fitted at both Tier 2 ($T = 600$ strides, multi-bout) and Tier 1 ($T = 1200$, single bout). Estimated marginal means were evaluated at normal walking speed (speed_c = 0). For the Terrier dataset, we collapsed the auditory and visual cueing conditions into a single "cued" level after verifying modality equivalence on all outcomes via paired $t$-tests (all $p > 0.05$ for stride time); the primary model was thus a two-level cueing factor (uncued vs. cued), with the three-level (N/A/V) model retained in the Zenodo reproduction archive. For the Roerdink dataset, stride speed models included a pacing × walking area interaction (both centered), while stride time models (restricted to the paced conditions) included walking area as a single fixed effect.

Pairwise condition contrasts were obtained with the emmeans package using the "revpairwise" method [56]. To adjust for the correlation among the eight outcomes without the conservatism of a full Bonferroni correction (which assumes independence), we report 99% confidence intervals throughout. Degrees of freedom for LMMs were approximated via the Satterthwaite method; beta GLMMs used Wald $z$-statistics. Analyses were run in R [57,58] using lme4 [59], lmerTest [60], glmmTMB [61], and emmeans [56]. The analysis scripts were developed collaboratively with Claude Code (Anthropic) under the author's direction and supervision. Outputs were manually inspected at each pipeline stage, internal consistency between summary tables and raw model dumps was verified. All implementation decisions and result interpretations are the author's responsibility. The full scripts, intermediate outputs, raw model dumps, and a pinned R session description are deposited in the reproduction archive.

## 4. Results

The eight-model ARFIMA family was fitted to 698 time series across the three datasets (216 Terrier, 266 Hausdorff, 216 Roerdink), yielding 6,448 individual model fits with zero convergence failures. Boundary trapping of the fractional parameter ($|d|$ at ±0.5 with near-zero standard error), a known artifact of dynamic mean estimation in the arfima package [33,53], occurred in fewer than 0.3% of the 3,224 long-memory model fits (4 instances across all datasets), confirming that the $dmean = FALSE$ setting eliminated this problem.

Residual adequacy was assessed with Ljung–Box tests (lag = 10) applied to the residuals of the BIC-best model for each series. The two treadmill datasets showed near-nominal rejection rates: 4.6% for Terrier stride time and 4.9% for Roerdink stride speed, close to the expected 5% false-positive rate.

The overground Hausdorff dataset showed an elevated rate of 18.8%, concentrated in the metronomic conditions (26.6%) and rising with series length. The $d_{BMA}$ estimates, which average over all eight models weighted by BIC, are robust to incomplete residual whitening by any single best model.

### 4.1. Cueing transforms stride time from persistent to anti-persistent long memory

Isochronous cueing reversed the sign of the fractional differencing parameter ($d$) in stride time in both datasets (Table 2; Fig. 3, row 1). In the Terrier data, uncued treadmill walking produced weakly persistent dynamics ($d_{BMA}$ = +0.13, 99% CI [0.03, 0.23]), while cueing drove them to strongly anti-persistent values (−0.52 [−0.60, −0.44]), a shift of Δ = −0.65 ($t$ = −14.82, $p$ < 0.0001). Auditory and visual cueing produced statistically indistinguishable effects (A: −0.54; V: −0.50; difference = 0.04, $p$ = 0.74), justifying their collapse into a single cued level. The Hausdorff data replicated this result in overground walking. The pacing-induced $d_{BMA}$ shift was −0.65 at $L$ = 600 ($t$ = −20.91, $p$ < 0.0001) and −0.70 at $L$ = 1200 ($t$ = −9.77, $p$ < 0.0001), closely matching the Terrier estimate (Table 2). Metronomic $d_{BMA}$ stabilized near −0.47 to −0.48 regardless of dataset, walking environment, or series length. Walking speed had no significant main effect or interaction with pacing on $d_{BMA}$ in either tier (all $p$ > 0.14).

DFA scaling exponents confirmed the same qualitative pattern (Table 2). Unconstrained walking yielded $\alpha$ ≈ 0.91–0.98 in all datasets and tiers; cueing collapsed $\alpha$ to 0.30–0.35 (Terrier: Δ = −0.61; Hausdorff T2: Δ = −0.60; both $p$ < 0.0001). Both estimators converge: cueing replaces persistence with anti-persistence.

**Table 2.** ARFIMA and DFA outcomes for stride time series. Estimated marginal means with 99% confidence intervals.

| | **Terrier (2016)**<br>**N = 36, T = 500** | | | **Hausdorff (1996) — T2**<br>**N = 10, L = 600** | | | **Hausdorff (1996) — T1**<br>**N = 10, L = 1200** | | |
|---|---|---|---|---|---|---|---|---|---|
| **Outcome** | **Uncued** | **Cued** | **Δ / OR** | **Unconstr.** | **Metro.** | **Δ / OR** | **Unconstr.** | **Metro.** | **Δ / OR** |
| $d_{BMA}$ (LMM) | +0.13<br>[0.03, 0.23] | −0.52<br>[−0.60, −0.44] | Δ = −0.65*** | +0.17<br>[0.11, 0.23] | −0.48<br>[−0.56, −0.41] | Δ = −0.65*** | +0.23<br>[0.09, 0.37] | −0.47<br>[−0.60, −0.33] | Δ = −0.70*** |
| $DFA\ \alpha$ (LMM) | 0.91<br>[0.84, 0.97] | 0.30<br>[0.24, 0.35] | Δ = −0.61*** | 0.95<br>[0.90, 1.00] | 0.35<br>[0.29, 0.41] | Δ = −0.60*** | 0.98<br>[0.90, 1.05] | 0.33<br>[0.25, 0.41] | Δ = −0.65*** |
| $W_{long}$ (beta) | 0.63<br>[0.51, 0.75] | 0.83<br>[0.76, 0.88] | OR = 2.74*** | 0.66<br>[0.59, 0.73] | 0.81<br>[0.73, 0.87] | OR = 2.15*** | 0.74<br>[0.60, 0.84] | 0.79<br>[0.66, 0.88] | OR = 1.32 ns |
| $W_{FI}$ (beta) | 0.21<br>[0.13, 0.31] | 0.10<br>[0.06, 0.15] | OR = 0.41*** | 0.16<br>[0.11, 0.23] | 0.09<br>[0.05, 0.15] | OR = 0.51*** | 0.11<br>[0.06, 0.20] | 0.09<br>[0.05, 0.17] | OR = 0.82 ns |
| $W_{ARFI}$ (beta) | 0.24<br>[0.15, 0.35] | 0.34<br>[0.26, 0.43] | OR = 1.68 ns | 0.32<br>[0.24, 0.41] | 0.41<br>[0.30, 0.53] | OR = 1.49 ns | 0.39<br>[0.26, 0.55] | 0.38<br>[0.25, 0.54] | OR = 0.96 ns |
| $W_{FIMA}$ (beta) | 0.16<br>[0.10, 0.25] | 0.11<br>[0.07, 0.16] | OR = 0.64 ns | 0.11<br>[0.09, 0.14] | 0.06<br>[0.04, 0.09] | OR = 0.49*** | 0.13<br>[0.07, 0.23] | 0.09<br>[0.05, 0.17] | OR = 0.65 ns |

| | Terrier (2016) N = 36, T = 500 | | | Hausdorff (1996) — T2 N = 10, L = 600 | | | Hausdorff (1996) — T1 N = 10, L = 1200 | | |
|---|---|---|---|---|---|---|---|---|---|
| **Outcome** | **Uncued** | **Cued** | **Δ / OR** | **Unconstr.** | **Metro.** | **Δ / OR** | **Unconstr.** | **Metro.** | **Δ / OR** |
| $W_{ARMA}$ (beta) | 0.31 [0.21, 0.43] | 0.11 [0.07, 0.16] | OR = 0.26*** | 0.32 [0.26, 0.39] | 0.14 [0.09, 0.20] | OR = 0.33*** | 0.23 [0.14, 0.36] | 0.16 [0.09, 0.27] | OR = 0.64 ns |
| $H_{norm}$ (LMM) | 0.40 [0.30, 0.49] | 0.32 [0.26, 0.39] | Δ = −0.07 ns | 0.41 [0.34, 0.47] | 0.26 [0.19, 0.34] | Δ = −0.14*** | 0.32 [0.20, 0.44] | 0.23 [0.11, 0.35] | Δ = −0.09 ns |

Outcomes: $d_{BMA}$, DFA α and $H_{norm}$ fitted with linear mixed models (LMM); $W_{long}$, $W_{FI}$,$W_{ARFI}$, $W_{ARFI}$ and $W_{ARMA}$fitted with beta GLMMs (logit link). All models include (1 | subject) random intercept.
Terrier conditions: Uncued = position-referenced treadmill walking; Cued = auditory and visual isochronous cueing collapsed
Hausdorff conditions: emmeans evaluated at normal walking speed (speed_c = 0), marginalized over slow and fast. T2 = L = 600 strides (multi-bout); T1 = L = 1200 strides (single bout).
Δ = cued − uncued (LMM outcomes, original scale). OR = odds ratio cued/uncued (beta outcomes, response scale). Significance: *** $p < 0.001$; * $p < 0.01$; ns = not significant. *p*-values from Satterthwaite *t*-tests (LMM outcomes) or Wald *z*-tests (beta GLMM outcomes).
99% confidence intervals (CI). CIs for LMM outcomes use Kenward–Roger degrees of freedom; CIs for beta outcomes are back-transformed from the logit scale.

#### *4.1.1. Long-memory models dominate in both persistent and anti-persistent conditions*

Whether anti-persistent stride time reflects genuine long-memory dynamics or short-memory ARMA processes is the central inferential question. The Schwarz weight evidence is unambiguous (Table 2). In both datasets, the aggregate long-memory weight $W_{long}$ was higher under cueing than under unconstrained walking: Terrier 0.63 uncued versus 0.83 cued (OR = 2.74, $p < 0.0001$); Hausdorff T2 0.66 versus 0.81 (OR = 2.15, $p < 0.0001$). Anti-persistent stride time series were thus *more* likely to be classified as long-memory than their persistent counterparts.

Model selection certainty, quantified by the normalized Shannon entropy $H_{norm}$ (0 = single model dominates; 1 = uniform weights), was moderate in all conditions (range: 0.23–0.41). Entropy decreased under cueing in the Hausdorff T2 data (Δ = −0.14, $p < 0.0001$), reflecting increasing concentration of weight on ARFI. In the Terrier data the decrease was smaller and non-significant (Δ = −0.07, $p = 0.095$).

#### *4.1.2. Series length modulates estimation precision but not the core finding*

The Hausdorff three-tier design provides a direct assessment of how series length affects ARFIMA inference. The persistent–anti-persistent contrast in $d_{BMA}$ was robust at both T1 and T2 lengths: Δ = −0.65 at $L = 600$ and Δ = −0.70 at $L = 1200$, with overlapping confidence intervals (Table 2; Fig. 3, column B vs. C).

The principal effect of series length appeared in the persistent conditions. Unconstrained walking produced $d_{BMA} = +0.17$ at $L = 600$ but +0.23 at $L = 1200$, and $W_{long}$ rose from 0.66 to 0.74. This asymmetry is expected: anti-persistent autocorrelations are concentrated in the first 10–15 lags and are captured even in short series, whereas persistent autocorrelations decay slowly as a power law and require longer series for the tail to emerge above sampling noise [33].

A consequence at T1 (L = 1200) was the loss of statistical significance in the model-weight contrasts between cueing conditions. All five weight outcomes were non-significant at T1 (Table 2, last column). At T2, four of these contrasts had been strongly significant; only $W_{ARFI}$ was already non-significant at T2 (Table 2). The pattern is interpretable: at T1, both conditions converge toward high long-memory classification ($W_{long} \approx$ 0.74 vs. 0.79), eliminating the differential that existed at T2 where short-memory ARMA models still competed effectively in the unconstrained condition. Because FI gains weight in unconstrained walking at longer series length (0.16 → 0.11) while ARFI remains the dominant long-memory model in both conditions, the within-long-memory redistribution also compresses. Longer series resolve model ambiguity in both directions, narrowing the between-condition gap in model selection even as the $d_{BMA}$ contrast grows slightly stronger.

A third tier extending the analysis to L = 2000 strides (Supplementary Table S1) was computed to test whether the T1 estimates are sensitive to further series-length extension. The four primary outcomes were essentially invariant between T1 and T3: $d_{BMA}$ moved from 0.230 [0.091, 0.368] to 0.252 [0.138, 0.367] (+9.9%), DFA α was unchanged at 0.977 (−0.1%), $\varphi_{BMA}$ moved from 0.253 to 0.220 (−13.1%), and $\theta_{BMA}$ was unchanged at −0.205 (+0.1%). All four 99% CIs at T3 overlapped broadly with the T1 values. What did change was the distribution of Schwarz weight: aggregate long-memory evidence $W_{long}$ rose from 0.740 to 0.836 (+13.0%) at the expense of the short-memory competitor $W_{ARMA}$ (0.232 → 0.164, −29.4%), and within the long-memory family the simpler FI parameterisation lost weight to the richer FIMA and ARFIMA specifications, with ARFI unchanged. These shifts follow the classical BIC behavior: as N grows, the likelihood gain of correctly specified richer models dominates the log N penalty [40]. The essential inference is that the T1 parameter estimates already reported in Table 2 are robust to substantial series-length extension; longer series sharpen model-selection evidence without altering the ARFIMA/DFA estimates themselves.

#### *4.1.3. Cueing reconfigures the short-memory parameter structure*

The sign reversal in *d* was accompanied by a reorganization of both the model selection landscape and the short-memory ARFIMA components (Table 2; Fig. 3, rows 3–4).

Under unconstrained walking, BIC evidence was distributed across competing models. In the Terrier data, $W_{ARMA}$ = 0.31 and $W_{ARFI}$ = 0.24, with $W_{FI}$ (0.21) and $W_{FIMA}$ (0.16) each receiving non-trivial weight; the full ARFIMA model contributed only modestly $W_{ARFIMA} \approx 0.02$. A similar pattern appeared in the Hausdorff T2 data, where ARFI (0.32) and ARMA (0.32) were nearly tied, with FI (0.16), FIMA (0.11), and ARFIMA (≈ 0.07) sharing the remainder. This dispersed weight distribution reflects genuine model selection ambiguity: with weakly persistent dynamics ($d_{BMA} \approx$ +0.13 to +0.17), the likelihood surface offers comparably good fits through different parameter trade-offs: pure long memory (FI), short memory with an AR correction (ARMA), or mixed configurations balancing $d$, φ, and θ (ARFI, FIMA, ARFIMA). At T1 (L = 1200), longer series began to resolve this ambiguity: $W_{ARFI}$ rose to 0.39 while $W_{ARMA}$ retreated to 0.23, consistent with the emergence of genuine long-memory structure above sampling noise.

Cueing produced a markedly different landscape. Long-term models containing a $\varphi$ parameter (i.e., ARFI and ARFIMA combined) captured 0.62 (Terrier), 0.66 (Hausdorff T2), and 0.61 (Hausdorff T1) of total weight, whereas FI dropped to ≈ 0.09–0.10 and FIMA to 0.06–0.11 (Table 2). Two factors underlie this rapid convergence. First, anti-persistent autocorrelations are concentrated in the first 10–15 lags and are detectable even in short series [33], reducing the estimation uncertainty that plagued the weakly persistent condition. Second, the correction mechanism imposed by the metronome generates a strong autoregressive signature that is unambiguously captured by φ.

The BMA-weighted parameter estimates confirmed this interpretation (Fig. 3, rows 3–4). Under cueing, $\varphi_{BMA}$ shifted upward: from 0.36 to 0.59 in the Hausdorff T2 data ($\Delta = +0.22$, $t = 3.77$, $p = 0.0002$) and from 0.25 to 0.59 at T1 ($\Delta = +0.34$, $t = 3.41$, $p = 0.0013$). In the Terrier data, the increase was modest and non-significant (0.43 to 0.51, $\Delta = +0.08$, $p = 0.22$), likely reflecting greater estimation noise at T = 500. The complementary trajectory appeared in $\theta_{BMA}$: under unconstrained walking, $\theta_{BMA}$ was negative in both datasets (Terrier: −0.35; Hausdorff T2: −0.28), consistent with the Wing–Kristofferson motor delay (Section 2.6). Cueing eliminated this MA component. In the Terrier data, $\Delta = +0.47$ ($t = 9.33$, $p < 0.0001$); in the Hausdorff T2 data, $\Delta = +0.30$ ($t = 8.11$, $p < 0.0001$); the T1 estimates pointed in the same direction ($\Delta = +0.20$, $p = 0.0024$).

The weight and parameter evidence converge on a coherent picture. In unconstrained walking, the model-selection landscape is fragmented: a negative $\theta$ from motor delay differencing, residual short-memory structure captured by a moderate $\varphi$, and a positive $d$ from the fractal timekeeper combine in varying proportions across the model family, and BIC cannot clearly adjudicate at T ≤ 600. Cueing simplifies this landscape: the phase-correction loop produces a detectable AR signature (φ rises above the unconstrained baseline), the MA component collapses ($\theta_{BMA} \rightarrow 0$), and FI becomes inadequate, leaving ARFI as the dominant model structure.

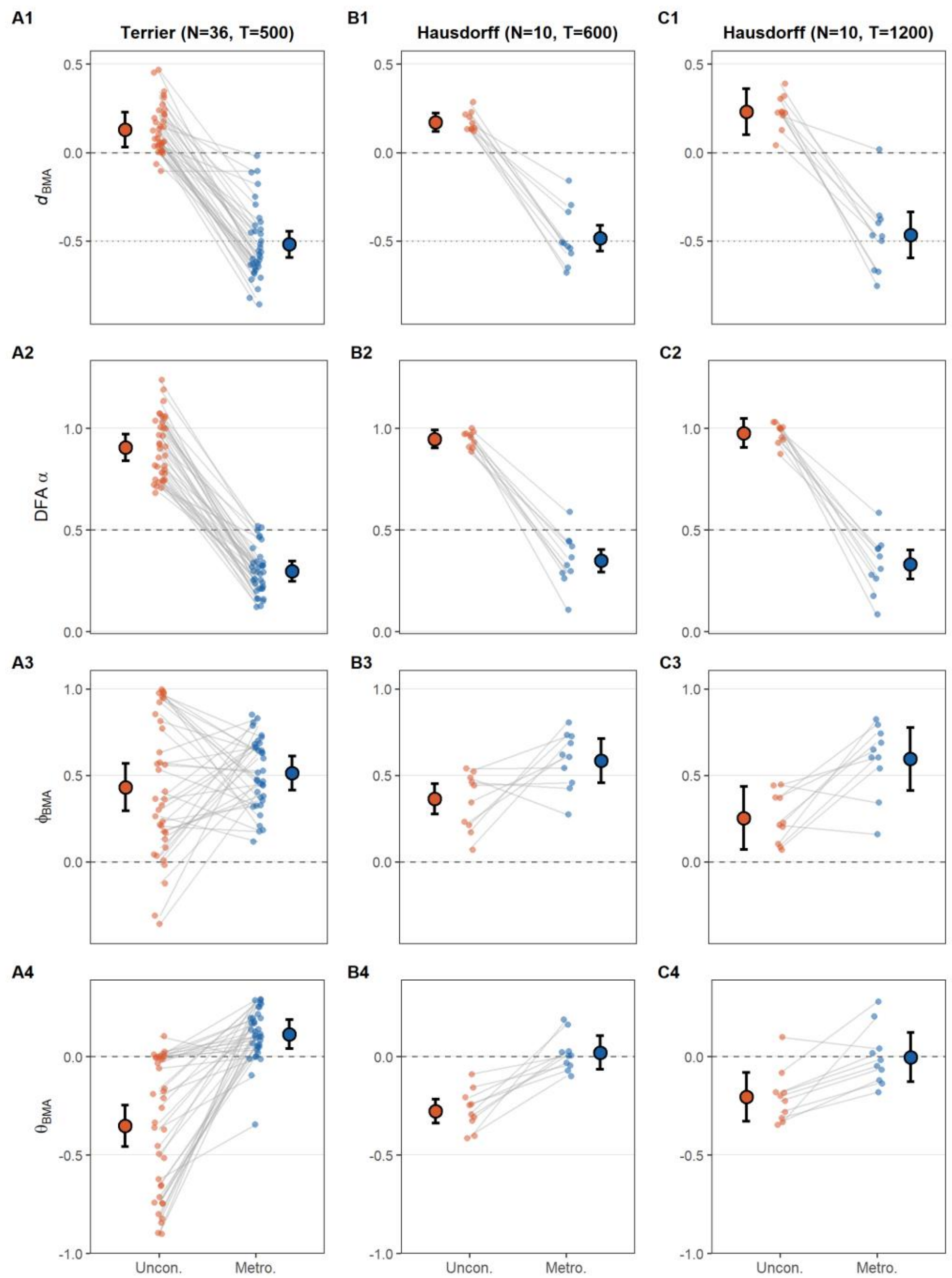

***Fig. 3. Effect of metronomic pacing on stride time ARFIMA and DFA parameters.***

*Rows show four parameters: $d_{BMA}$ (BMA-weighted fractional differencing parameter; A1–C1), DFA $\alpha$ (DFA scaling exponent; A2–C2), $\varphi_{BMA}$ (BMA-weighted autoregressive coefficient; A3–C3), and $\theta_{BMA}$ (BMA-weighted moving-average coefficient; A4–C4). Columns show three dataset–series length combinations: Terrier (N = 36, T = 500; A+V), Hausdorff Tier 2 (N = 10, T = 600; column B), and Hausdorff Tier 1 (N = 10, T = 1200; column C). Small dots are individual participants under unconstrained walking (coral) and metronomic pacing (blue); grey lines connect the same participant across conditions. For the Hausdorff dataset, each dot is the participant's mean across three walking speeds. Large circles with error bars are estimated marginal means (emmeans) and 99 % confidence intervals from linear mixed-effects models (emmean $\pm z_{0.995} \times SE$; $z_{0.995} = 2.576$). For the Terrier dataset, auditory and visual cueing were collapsed into a single cued level. Dashed lines mark white-noise reference values ($d_{BMA} = 0$, DFA $\alpha = 0.5$, $\varphi_{BMA} = 0$, $\theta_{BMA} = 0$); in row 1, the dotted line marks $d = -0.5$. ARFIMA: autoregressive fractionally integrated moving average; BMA: Bayesian model averaging (BIC Schwarz weights); DFA: detrended fluctuation analysis.*

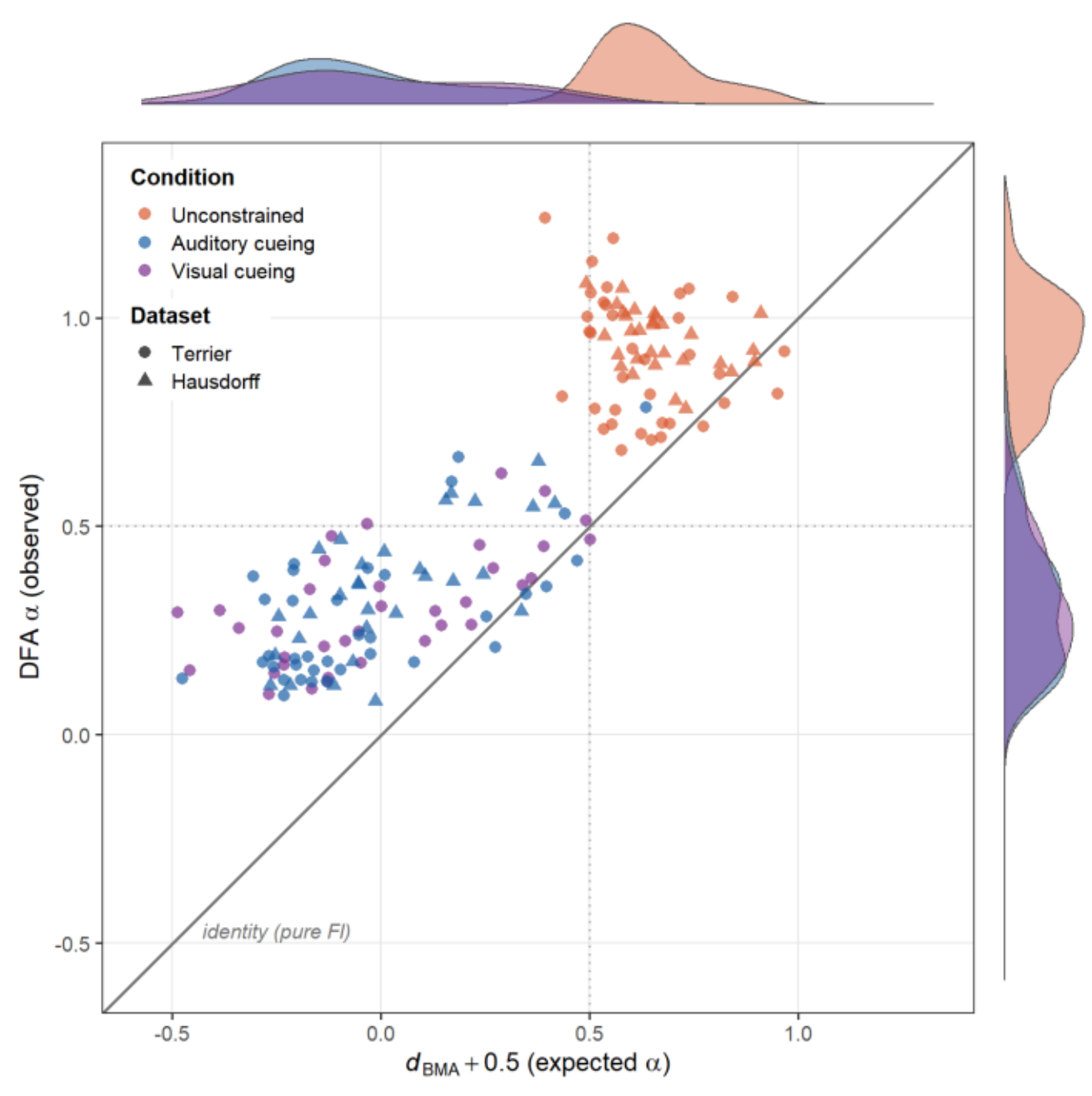

***Fig. 4. Observed DFA scaling exponent versus ARFIMA-predicted exponent for stride time.***
*Each point plots the observed DFA α against $d_{BMA}$ + 0.5, the value expected under a pure fractionally integrated (FI) process. Circles: Terrier dataset (N = 36, three conditions); triangles: Hausdorff Tier 2 dataset (N = 10, bout-averaged per speed × pacing combination). Colors distinguish unconstrained walking (coral), auditory/metronomic cueing (blue), and visual cueing (purple; Terrier only). The diagonal line is the identity α = d + 0.5 expected for pure FI processes; points above this line indicate that DFA overestimates the long-memory exponent. Dotted crosshairs mark α = 0.5 (white noise). Marginal density strips show the univariate distributions of each axis by condition. Total: 168 points (108 Terrier + 60 Hausdorff); participants contribute multiple points across conditions.*

### *4.1.4. DFA α exceeds ARFIMA d + 0.5 across all conditions*

For a pure fractionally integrated process, the DFA scaling exponent equals *d* + 0.5 (Eq. 21). This identity provides a direct cross-check between the two estimators: in the absence of short-memory contamination, DFA α and ARFIMA d + 0.5 should coincide. In our data, however, observed α exceeded this prediction in the large majority of series. Fig. 4 plots DFA α against d_BMA + 0.5 for all stride time series from both datasets. The unconstrained cluster sits above the identity line, with a mean positive bias of approximately +0.28 α units in both Terrier and Hausdorff T2. The cued/metronomic cluster also lies above identity, with a slightly larger bias of approximately +0.32 (Terrier) and +0.33 (Hausdorff T2). The bias was present at both series lengths in the Hausdorff data, indicating that it does not vanish as T increases. Two factors are expected to contribute. First, DFA α conflates short-memory and long-memory contributions when ARMA components coexist with the fractional differencing parameter, inflating α above d + 0.5. Second, the ML estimator of d carries a known finite-sample negative bias (Section 2.3), which deflates d_BMA + 0.5 below its asymptotic value.

## 4.2. Part II: Stride speed anti-persistence under graded constraint

On a constant-speed treadmill, stride speed is anti-persistent in every condition because the fixed belt forces active speed regulation (Section 1). The question addressed here is whether the *degree* of anti-persistence varies with constraint tightness, and whether the ARFIMA model structure changes across constraint types. Table 3 reports the parameter estimates for all SS conditions; Fig. 5 presents the model selection results.

**Table 3. Stride speed anti-persistence**

| | Fractal estimation | | | Correction parameters | |
|---|---|---|---|---|---|
| **Condition** | $d_{\mathrm{BMA}}$ | **DFA** $\alpha$ | $\Delta\alpha$ | $\varphi_{\mathrm{BMA}}$ | $\theta_{\mathrm{BMA}}$ |
| ***Terrier (2016)***<br>N = 36, T = 500 | | | | | |
| No cueing | −0.31<br>[−0.41, −0.21] | 0.30<br>[0.25, 0.35] | +0.12<br>[+0.03, +0.21] | +0.21<br>[+0.07, +0.35] | −0.08<br>[−0.16, −0.01] |
| Auditory cueing | −0.35<br>[−0.44, −0.26] | 0.30<br>[0.25, 0.35] | +0.14<br>[+0.05, +0.24] | +0.23<br>[+0.07, +0.40] | −0.07<br>[−0.18, +0.03] |
| Visual cueing | −0.40<br>[−0.52, −0.29] | 0.26<br>[0.19, 0.32] | +0.16<br>[+0.05, +0.27] | +0.36<br>[+0.21, +0.51] | −0.09<br>[−0.23, +0.06] |
| ***Roerdink (2019) — small area***<br>N = 24, T = 256 | | | | | |
| Unpaced | −0.33<br>[−0.45, −0.21] | 0.18<br>[0.12, 0.23] | +0.01<br>[−0.11, +0.13] | +0.10<br>[+0.00, +0.20] | −0.26<br>[−0.42, −0.10] |
| Paced | −0.39<br>[−0.49, −0.30] | 0.18<br>[0.12, 0.24] | +0.07<br>[−0.01, +0.15] | +0.07<br>[−0.02, +0.15] | −0.18<br>[−0.29, −0.08] |
| ***Roerdink (2019) — intermediate area***<br>N = 24, T = 256 | | | | | |
| Unpaced | −0.30<br>[−0.40, −0.20] | 0.24<br>[0.19, 0.30] | +0.05<br>[−0.02, +0.12] | −0.02<br>[−0.15, +0.10] | −0.09<br>[−0.19, +0.02] |
| Paced | −0.28<br>[−0.39, −0.17] | 0.24<br>[0.19, 0.30] | +0.02<br>[−0.07, +0.11] | +0.01<br>[−0.09, +0.11] | −0.14<br>[−0.26, −0.02] |
| ***Roerdink (2019) — large area***<br>N = 24, T = 256 | | | | | |
| Unpaced | −0.16<br>[−0.25, −0.08] | 0.34<br>[0.28, 0.40] | +0.00<br>[−0.06, +0.07] | +0.02<br>[−0.09, +0.13] | −0.15<br>[−0.25, −0.06] |
| Paced | −0.20<br>[−0.28, −0.11] | 0.33<br>[0.27, 0.40] | +0.03<br>[−0.03, +0.10] | +0.01<br>[−0.08, +0.09] | −0.12<br>[−0.20, −0.04] |

$d_{\mathrm{BMA}}$ = BMA-weighted fractional differencing parameter (Schwarz weights across 8 models; short-memory models contribute $d = 0$).
DFA $\alpha$ = scaling exponent. $\Delta\alpha$ = DFA $\alpha - (d_{\mathrm{BMA}} + 0.5)$: positive values indicate DFA overestimation.
$\varphi_{\mathrm{BMA}}$ = BMA-weighted autoregressive coefficient. $\theta_{\mathrm{BMA}}$ = BMA-weighted moving-average coefficient. Both computed as $\Sigma W_i \times$ parameter$_i$ across all 8 models (= 0 for models without the component).
Terrier: C-Mill treadmill. No cueing = position-referenced walking; Auditory/Visual = isochronous cueing. All SS series are anti-persistent due to fixed belt speed. Roerdink: C-Mill treadmill. Walking area constrains spatial displacement. Paced = isochronous metronome.
99% CIs.

The Roerdink walking area manipulation produced a clear gradient in $d_{BMA}$ (Table 3). Anti-persistence was strongest in the small walking area ($d_{BMA} \approx -0.33$ to $-0.39$) and weakest in the large area (−0.16 to −0.20; Table 3). The walking area slope was +0.090 per unit ($t$ = 5.73, $p$ < 0.0001), corresponding to an increase of approximately 0.18 $d_{BMA}$ units from the small to the large area. Even the least constrained configuration (large area, unpaced) remained significantly anti-persistent: $d_{BMA}$ = −0.16 [99% CI: −0.25, −0.08]. DFA α tracked the same gradient (slope = +0.080, $t$ = 11.05, $p$ < 0.0001).

Acoustic pacing had no significant effect on SS anti-persistence. In the Roerdink data, the pacing coefficient on $d_{BMA}$ was −0.025 ($t$ = −0.96, $p$ = 0.34), and none of the remaining seven outcomes reached significance (all $p$ > 0.24). The pacing × walking area interaction was likewise non-significant for all eight outcomes (all $p$ > 0.44).

In the Terrier data, SS was anti-persistent in all three conditions (Table 3). A trend toward stronger anti-persistence emerged from no cueing ($d_{BMA}$ = −0.31) through auditory cueing (−0.35) to visual cueing (−0.40), but no pairwise contrast reached significance (all $p$ > 0.17). DFA α showed the same ordering (0.30, 0.30, 0.26 for NC, AC, VC) without significant differences (all $p$ > 0.10).

#### *4.2.1. FI dominates stride speed; visual cueing promotes ARFI*

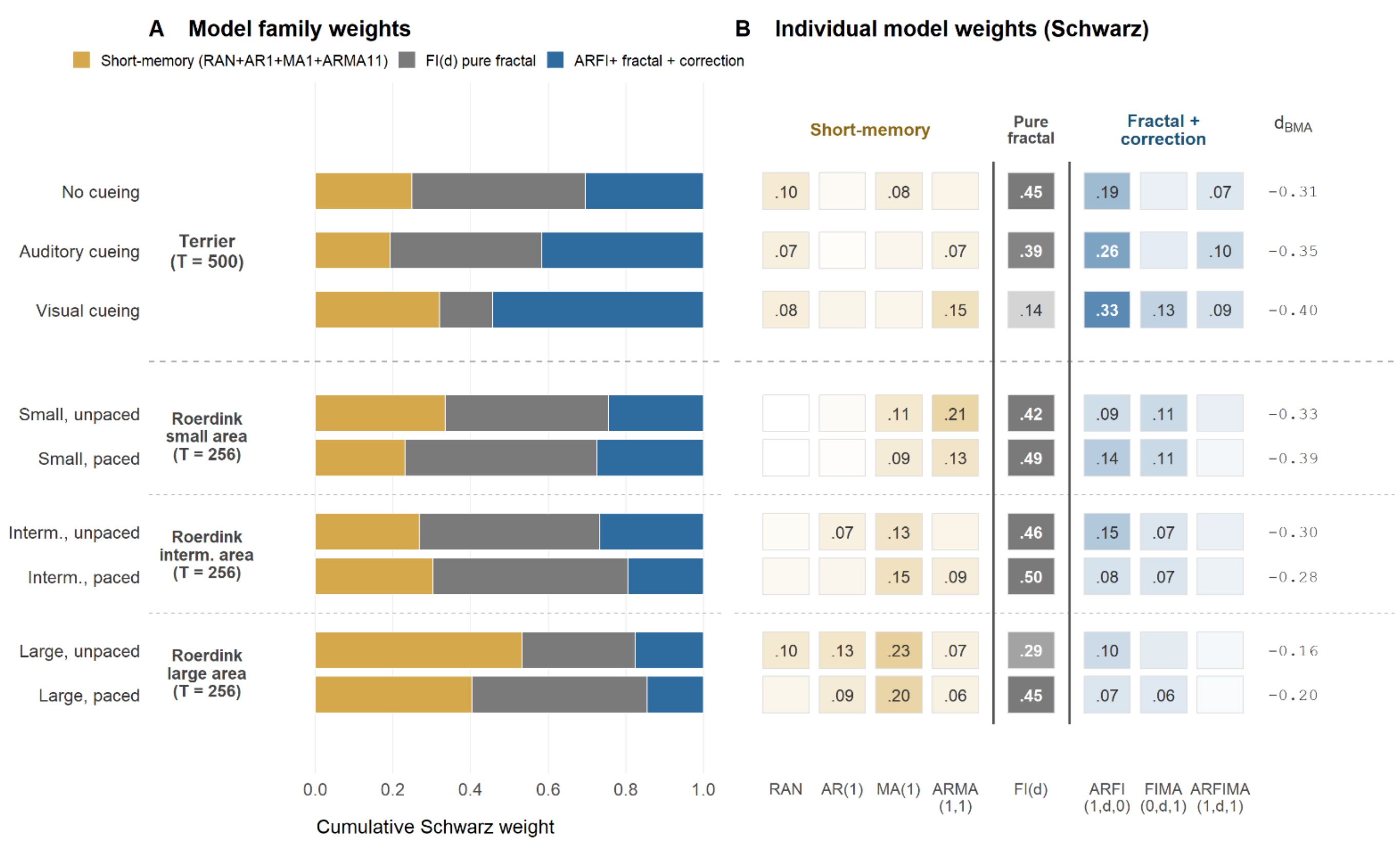

***Fig. 5. ARFIMA model selection for stride speed across two datasets.*** *Nine conditions are shown: three from the Terrier dataset (T = 500 strides; no cueing, auditory cueing, visual cueing) and six from the Roerdink dataset (T = 256; three walking areas × two pacing levels).* ***(A)*** *Stacked bars show cumulative BIC-derived Schwarz weights grouped into three model families: short-memory (RAN + AR(1) + MA(1) + ARMA(1,1); gold), pure fractional (FI(d); grey), and fractal-with-correction (ARFI + FIMA + ARFIMA; blue). Each bar sums to 1.0.* ***(B)*** *Heatmap of individual Schwarz weights for all eight candidate models; color intensity scales with weight (darker = higher). Values ≥ 0.055 are printed; values ≥ 0.25 are bolded. The rightmost column reports the group-*

*mean $d_{BMA}$. All weights are group means: N = 36 participants (Terrier) and N = 24 participants (Roerdink). Dashed lines separate datasets; within the Roerdink dataset, conditions are grouped by walking area.*

Long-memory models captured the majority of Schwarz weight in all conditions ($W_{long}$ = 0.47–0.81; Fig. 5, Panel A), confirming that SS anti-persistence is a genuine fractal phenomenon across both datasets. The dominant long-memory model, however, differed from the ST results. Whereas cueing shifted the dominant ST model toward ARFI(1,$d$,0) (Section 4.1.1), the pure fractional FI($d$) specification dominated SS in the majority of conditions (Fig. 5, Panel B).

The Terrier data revealed a condition-dependent shift. Under no cueing, FI carried the largest individual weight (0.45) and ARFI was secondary (0.19). Auditory cueing redistributed weight modestly toward ARFI (0.26 vs. FI = 0.39). Visual cueing reversed the dominance entirely: ARFI became the leading model (0.33) and FI dropped to 0.14. This FI-to-ARFI transition under visual cueing coincides with the documented increase in fluctuation magnitude specific to visual cueing reported for this dataset [13]. The short-memory ARMA model, which competed effectively in unconstrained ST ($W_{ARMA} \approx 0.31$; Section 4.1.1), was suppressed in SS throughout (0.05–0.15).

In the Roerdink data, model selection was less differentiated. FI dominated in every condition (0.29–0.50), and the remaining weight was distributed broadly across several models (Fig. 5, Panel B). In the large walking area (unpaced), where $d_{BMA}$ approached zero (−0.16), the short-memory MA(1) model reached 0.23, approaching the FI weight of 0.29; the aggregate $W_{long}$ dropped to 0.47. This dilution of model selection certainty is expected when the long-memory signal weakens: as $d_{BMA}$ approaches the short-memory/long-memory boundary, competing models become harder to discriminate. Beyond the robust FI dominance and the overall long-memory classification, finer distinctions in model structure are unreliable at T = 256.

#### 4.2.2. The DFA – ARFIMA gap is smaller for stride speed

The discrepancy between DFA α and $d_{BMA}$ + 0.5 was smaller for SS than for ST. In the Terrier data, the mean Δα ranged from +0.12 to +0.16 across conditions, compared with +0.28 for ST (Section 4.1.4). In the Roerdink data, $\Delta\alpha$ was near zero in most conditions (range: 0.00 to +0.07; Table 3). The reduced bias is consistent with the dominance of the pure FI model: when short-memory components carry little weight, DFA α approximates $d_{BMA}$ + 0.5 as predicted by Eq. (21). The Terrier data, where ARFI carried more weight (particularly under visual cueing), showed correspondingly larger $\Delta\alpha$.

#### 4.2.3. Short-memory parameter estimation requires T ≥ 500

The short-memory parameters showed a striking dissociation between datasets (Table 3). In the Terrier data (T = 500), $\varphi_{BMA}$ was the dominant short-memory coefficient: +0.21 to +0.36 across conditions, with the 99% CI excluding zero in all three. The $\theta_{BMA}$ estimates were small (+0.07 to +0.09) and the CI included zero in two of three conditions. The pattern reversed completely in the Roerdink data (T = 256): $\theta_{BMA}$ dominated (+0.12 to +0.26, CI excluding zero in five of six conditions) while $\varphi_{BMA}$ collapsed to near zero (−0.02 to +0.10, CI including zero in all six conditions).

This φ/θ reversal occurred despite both datasets being collected on the same instrumented treadmill (C-Mill) under comparable belt speed constraints. We attribute it to a series-length artifact rather than a genuine mechanistic difference. At T = 256, the sample autocorrelation function carries insufficient precision beyond the first few lags to discriminate between an AR(1) process (exponentially decaying ACF) and an MA(1) process (single lag-1 spike). Because both produce similar lag-1 autocorrelation, the maximum likelihood estimator defaults to the more parsimonious MA(1) representation. This interpretation is supported by Braun [34], who demonstrated that reliable AR/MA discrimination in ARFIMA models requires T ≥ 500.

# 5. Discussion

## 5.1. Anti-persistence as a rare fractal phenomenon

Genuine anti-persistent long-range dependence is exceedingly rare. Hydrological records, precipitation proxies, and paleoclimate series are uniformly persistent [62]; apparent anti-persistence in financial returns traces to improper detrending rather than any true fractal structure [63]; ARFIMA-GARCH modeling of long-term heart rate recordings reveals persistent long memory in normal sinus rhythm, with no evidence of anti-persistent dynamics even in pathological conditions such as heart failure or atrial fibrillation [64]. Similarly, interbreath interval time series in healthy adults show persistent long-range correlations with no reliable reports of anti-persistence even in pathology [65,66]. Neural oscillations [67] and DNA sequences [2] likewise fall on the persistent or memoryless side of the spectrum. The few confirmed exceptions are confined to rhythmic motor control: metronomic finger tapping [24,25,68,69], postural velocity during quiet standing [70], short-timescale inter-beat intervals in expert drumming [71], and, as our data confirm, cued gait.

This specificity is not coincidental. As established in Section 2.1, the spectral zero at $f = 0$ that defines anti-persistent long memory requires the process variance (the zero-lag autocovariance, positive by definition) to be exactly cancelled by the cumulative autocovariances at all nonzero lags. Granger's aggregation theorem [30] shows that hierarchical multi-timescale systems naturally produce persistent long memory ($d > 0$); the spectral zero required by $d < 0$ arises only when active error correction against an external reference drives the autocovariance structure toward this precise cancellation. What distinguishes metronomic cueing from unconstrained walking, and fixed-speed treadmill walking from self-paced locomotion, is the presence and strength of exactly that correction.

## 5.2. A unified error-correction architecture

The results invite a single interpretation rather than a catalog of special cases. We argue that the architecture introduced in Section 2.6 (a fractal timekeeper, a reactive correction against an external reference, and a shared motor delay at stride boundaries) operates across the anti-persistent conditions we observed, with the nature of the reference and the gain of the correction varying from one context to another. The subsections that follow develop this claim along a gradient of decreasing external constraint: rhythmic cueing of stride time, where the correction is stride-locked and strongly expressed (section 5.2.1) and stride speed on a fixed-speed belt, where the reference is a velocity rather than a beat and the correction is weaker and more diffuse (section 5.2.2).

### *5.2.1. Rhythmic cueing and stride time*

Under isochronous cueing, stride time $d_{BMA}$ ranged from −0.47 to −0.52 in both the Terrier and Hausdorff datasets, with long-memory Schwarz weights of 0.81–0.83 against 0.11–0.14 for ARMA (Table 2). These numbers carry a weight that DFA scaling exponents alone cannot. A DFA $\alpha$ below 0.5 is consistent with anti-persistent long memory, but it is equally consistent with a short-memory ARMA process that contains no fractional component whatsoever: an ARMA(1,1) with a moderately strong autoregressive coefficient and a negative moving-average coefficient can produce the same slow power-law-like decay in the log-log fluctuation plot that DFA interprets as anti-persistence [17,18]. The entire cued-gait literature from Hausdorff et al. [1] onward has relied on DFA and therefore lacks a direct test of whether the observed correlation structure requires a fractional differencing parameter. Delignières and Torre [9] took a first step by fitting ARFIMA models to the original Hausdorff data, reporting negative $d$ estimates ($d \approx -0.33$ to $-0.44$) under metronomic conditions and identifying long-memory models as BIC-best. Their analysis confirmed the anti-persistent pattern under parametric estimation but stopped short of a formal model-family comparison. The present results close that gap. The balance of evidence is unambiguous: anti-persistence of stride time under cueing is a genuine fractional phenomenon, not an ARMA artifact.

The sensorimotor synchronization model of Torre and Delignières [25] was designed for periodic tasks. A fractal timekeeper generates persistent interval fluctuations $d_t$; at each stride, the walker computes an asynchrony with respect to the metronome beat and applies a proportional phase correction (see Section 2.6). The full theoretical prediction is therefore ARFIMA(1, $d$, 1) with $d < 0$ at the output (forced by the bounded-asynchrony), a positive autoregressive coefficient $\varphi > 0$ reflecting the residual short-memory structure of the correction loop, and a moving-average coefficient $\theta$ combining the Wing–Kristofferson motor delay with the correction residual. In practice, BIC penalizes the extra MA parameter when motor delay variance is small relative to timekeeper variance, and selects the more parsimonious ARFI(1, $d$, 0). This is what our data show: ARFI dominates BMA model selection under cueing in both Terrier and Hausdorff datasets (Table 2, Fig. 3), with $\varphi$ values consistent with an active correction loop. The precise mapping from $\hat{\varphi}$ to the underlying correction gain $\kappa$ is, however, non-trivial: under the bare Vorberg–Schulze model (Eq. 23), the AR coefficient of the inter-response intervals equals $(1-\kappa)$; under the Torre fractal extension (Eqs. 24–25), this relationship is biased downward by the spectral mixing of the long-memory generator with the short-memory correction loop [25].

#### *5.2.2. Speed regulation on a fixed treadmill: positional error correction without a beat*

Stride speed on a fixed-speed treadmill is anti-persistent in every condition we tested, including the absence of any rhythmic cueing (Table 3, Fig. 5). The BMA-weighted AR coefficient is positive in all three Terrier conditions ($\varphi_{BMA}$ = +0.21 to +0.36), reproducing the positive-$\varphi$ signature observed for cued ST. The MA coefficient is small ($\theta_{BMA}$ = −0.07 to −0.09) and mostly not distinguishable from zero, a pattern that mirrors what we observed for cued ST (Section 4.1.3). What differs is the magnitude. The AR coefficient for SS reaches +0.36 only under visual cueing, substantially lower than the values obtained for cued ST (Fig. 3). The presence of a positive AR signature is qualitatively consistent with an active correction loop in both paradigms, though the structural differences between phase correction against a metronome and positional correction against a fixed belt preclude a direct numerical comparison of the two AR coefficients in terms of correction strength.

This interpretation is supported by the condition-dependent model selection in the Terrier data. Under no cueing, FI dominates ($W_{FI}$ = 0.45, $W_{ARFI}$ = 0.19) because BMA cannot justify the extra AR parameter when the correction signal is weak relative to the series-length penalty. Auditory cueing produces a modest shift toward ARFI ($W_{ARFI}$ = 0.26). Visual cueing, which demands stride-by-stride spatial regulation against projected stepping stones, reverses the dominance: ARFI carries the leading weight (0.33) and FI drops to 0.14. At $T$ = 256 (Roerdink), the separation of AR and MA structure falls below the reliable threshold (Section 4.2.1), and finer model-structure distinctions become uninformative regardless of the underlying mechanism.

Why does the AR signature emerge less prominently for speed regulation than for rhythmic synchronization? Three structural features of the positional-control task may contribute. First, the belt imposes a speed rather than a rhythm, and there is no discrete timing reference against which each stride generates an asynchrony; the error variable is not defined at each event. Second, the relevant error is positional drift, an integrated quantity: stride-to-stride speed deviations accumulate as displacement, and the walker can tolerate moderate drift before reaching the uncomfortable edges of the belt. Third, the correction may operate intermittently rather than proportionally at every stride, with brief compensatory accelerations triggered when drift approaches a perceptual threshold. A weaker, more temporally diffuse correction signal of this kind would be harder to separate from the fractal component at available series lengths, pushing the Schwarz weight toward FI when the constraint is light and allowing ARFI to emerge only when the constraint tightens, as the Terrier gradient shows.

These features do not eliminate the correction; they attenuate it and blur its temporal fingerprint. Decker, Cignetti, and Stergiou [16] provided direct evidence that speed regulation is both real and cognitively mediated: a concurrent executive task weakened SS anti-persistence ($\alpha$ rising from 0.31 to 0.45), demonstrating that the correction draws on cortical attentional resources. In the present framework, diverting executive resources reduces the correction gain, $d$ moves toward zero, and $\alpha$ increases. That $\alpha$ remained below 0.5 even under cognitive load indicates the correction was weakened, not abolished, consistent with a graded gain rather than an on/off switch.

#### *5.2.3. Unconstrained walking and the residual correction*

The unified architecture makes a prediction about the third condition in our data: unconstrained walking, where neither a metronome nor a tight spatial constraint is imposed. Within the Torre framework, free overground walking corresponds to the limit $\kappa \to 0$ (no metronome reference, no asynchrony to correct), in which the fractal timekeeper alone shapes the inter-response intervals and the prediction reduces to FIMA(0, d, 1) with $d > 0$. The pure fractal limit is rarely realised in practice, however, because biomechanical references (preferred cadence, dynamic balance, metabolic optimum) may impose weak closed-loop regulation that injects residual short-memory structure into the observed series even in the absence of an explicit external pacing signal [25]. The Hausdorff T3 reanalysis at L = 2000 strides (Supplementary Table S1) provides the decisive test, for two converging reasons. First, the short-memory alternative (that unconstrained stride time reflects an ARMA regulator with no fractal component) is strongly disfavored: at T = 2000, the long-memory family carries more than five times the Schwarz weight of ARMA (0.84 vs. 0.16). Second, long-memory weight is distributed across ARFI (0.39), FIMA (0.16), and ARFIMA (0.31) together rather than concentrated on any one specification (see supplementary files S1), which points out cohabitation of AR and MA components in addition to the fractal differentiation parameter.

A complete validation, however, requires moving from the observed stride-interval parameters to the underlying mechanism the Torre architecture operates on, which is not the ITI itself but the asynchrony series, on which the fractal timekeeper, the motor-delay noise, and the correction toward $\tau$ act, and whose regulation produces the observed d deflation at the ITI output, as shown in Eqs. (18) and (19). Testing this claim requires running the architecture forward with candidate values of the timekeeper d, the motor-delay variance, and $\kappa$, analyzing the simulated ITI series with the same eight-model ARFIMA / BMA pipeline, and checking whether the recovered ARFIMA parameters $(d, \varphi, \theta)$ matches the empirical ranges reported here (Table 2, Figure 3 and supplementary material S1).

What reference would the walker be correcting against in the absence of a metronome or explicit spatial constraint? Several candidates exist. Overground, walkers maintain a roughly constant speed, follow a path, and preserve dynamic balance, all of which generate proprioceptive and visual error signals that feed back into stride timing. Metabolic optimization may impose a further constraint: stride intervals that deviate from the energetically optimal cadence incur a cost, and the locomotor system corrects toward the minimum [72]. None of these references is as discrete or as powerful as a metronome beat, which is precisely why the correction gain remains low and $d$ stays positive. The unified model does not require that the reference be external or that the walker be conscious of correcting; it requires only that some closed-loop regulation operate on the gait parameters, however weakly.

## 5.3. Methodological implications

The two methodological findings of the paper bear directly on how future gait-variability studies should be designed and interpreted. The first concerns the relationship between DFA α and the fractional differencing parameter; the second concerns the minimum series length required for reliable ARFIMA inference.

### *5.3.1. DFA α conflates long and short memory*

Across all three datasets, DFA $\alpha$ exceeded the value of d + 0.5 expected for a pure fractionally integrated process by a systematic margin of +0.25 to +0.34 $\alpha$ units, as shown in sections 4.1.4 and 4.2.2. Under persistent conditions, the bias is attributable to the positive autoregressive coefficient ($\varphi$ ≈ 0.2–0.4 in unconstrained stride time), which inflates DFA $\alpha$ by adding short-memory positive correlations at short lags that DFA cannot separate from long-memory persistence. Under anti-persistent conditions, a negative $\theta$ would be expected to deflate DFA $\alpha$, but the cued conditions show almost no $\theta$ ($\theta_{BMA}$ near zero in both Terrier and Hausdorff), leaving the AR inflation uncompensated. In unconstrained walking, $\varphi$ inflates and $\theta \approx -0.3$ deflates, and the two effects partially cancel; in cued walking the cancellation is lost and the bias is at its largest. A complementary contribution to the observed gap comes from the ML estimator of d itself: as established in Section 2.3, the finite-sample bias of exact ML deflates $\hat{d}$ in the persistent regime and amplifies the apparent excess of DFA α over $\hat{d} + 0.5$.

Heneghan and McDarby [73] derived an analytical equivalence between the DFA fluctuation function and a frequency-weighted integral of the power spectral density, implying that any non-fractal feature in the underlying process, including the smooth low-frequency deviations introduced by short-memory ARMA components, propagates directly into the estimated DFA exponent. Empirical investigations have repeatedly confirmed that DFA produces systematic biases when short-memory dynamics coexist with fractional integration, with the magnitude depending on the AR and MA coefficients [18,74,75]. Combined with the finite-sample negative bias of the ML estimator of d (Section 2.3), these two sources of distortion jointly account for the observed gap between DFA α and $\hat{d}_{BMA} + 0.5$.

#### *5.3.2. Series length constrains what can be estimated*

The three datasets span series lengths from T = 256 (Roerdink) to T = 2000 (Hausdorff T3), covering the full range typical of gait-variability studies. The core finding is that the long-memory family is detectable across the entire range: the fractional differencing parameter $d_{BMA}$ is robust, and the classification of a series as long-memory versus ARMA dominant is stable from T = 256 upward. What changes with T is the resolvability of model structure within the long-memory family. At T = 256, the autoregressive and moving-average components become statistically indistinguishable, and the $\varphi/\theta$ dominance pattern observed at T = 500 (AR-dominant under the same C-Mill protocol in the Terrier data) reverses at T = 256 despite the shared experimental platform. At T = 500–600, AR and MA are discriminable but the model-weight contrasts in persistent conditions remain underpowered: in the Hausdorff T2 data (L = 600), the ARFI-vs-FI difference in unconstrained walking fails to reach significance, and only in the T1 (L = 1200) reanalysis does the positive φ signature of a residual correction become statistically visible (see section 4.1.2). Practical recommendations follow: for tests of the sign of d —the persistence-versus-anti-persistence question — T ≥ 256 suffices; for the estimation of short-range parameters and for finer discrimination within the long-memory family, T ≥ 500 is advisable, and T ≥ 1000 is needed to detect weak AR components in near-persistent and persistent conditions.

## 6. Conclusion

We applied systematic ARFIMA model comparison to stride-to-stride variability in three independent gait datasets, estimating the fractional differencing parameter d and the autoregressive and moving-average coefficients $\varphi$ and $\theta$ under Bayesian model averaging across eight nested specifications. Three findings anchor the paper. First, anti-persistent long-range dependence under cueing and under fixed-speed treadmill walking is a genuine fractional phenomenon rather than a short-memory ARMA artifact: long-memory models prevail over ARMA(1,1) in every condition tested. Second, the sensorimotor synchronization architecture introduced by Torre and Delignières for rhythmic tapping extends to gait across a range of constraint types : rhythmic cueing of stride time, non-rhythmic positional constraint of stride speed on a constant belt, and likely unconstrained walking with an attenuated residual correction. Third, systematic comparison of $d_{BMA}$with DFA $\alpha$ quantifies a positive bias of +0.25 to +0.34 $\alpha$ units arising from short-memory components that DFA conflates with long-memory persistence, establishing ARFIMA-based decomposition as the more informative estimator for gait-variability research. Taken together, these results recast the catalog of persistent and anti-persistent findings in the gait-fractal literature as the visible output of a common closed-loop control architecture operating against external references of varying type and tightness, and provide the empirical baseline against which future mechanistic simulations of the timekeeper, correction loop, and motor-delay components can be tested.

**Declaration of competing interest**

The author declares no competing financial or non-financial interests relevant to the content of this article.

**Supplementary material**

Two supplementary documents are provided with this article: S1 — Effect of series length on ARFIMA and DFA outcomes (Hausdorff 1996, Tier 3 at L = 2000 strides), referenced in Sections 3.1 and 4.1.3; S2 — Morris elementary-effects screening of the ARFIMA/DFA pipeline, referenced in Section 3.4.

**CRediT authorship contribution statement**

Philippe Terrier: Conceptualization, Methodology, Software, Validation, Formal analysis, Investigation, Data curation, Writing – original draft, Writing – review & editing, Visualization, Project administration.

**Funding**

This research received no specific grant from any funding agency in the public, commercial, or not-for-profit sectors.

**Acknowledgments**

Institutional support was provided by Haute École Arc Santé, HES-SO University of Applied Sciences and Arts Western Switzerland.

**Data availability**

The analysis code, derived results, raw model output, validation figures, and the Terrier (2016) stride series redistributed under Creative Commons Attribution 4.0 International are openly available in the "Zenodo reproduction archive Behind the scenes of an ARFIMA gait study" at https://doi.org/10.5281/zenodo.19676064. The Hausdorff (1996) raw gait recordings are available from the PhysioNet umwdb collection (https://doi.org/10.13026/C28679) under the Open Data Commons Attribution Licence v1.0. The Roerdink et al. (2019) raw stride series are available in the supplementary material of the original publication (https://doi.org/10.3389/fphys.2019.00257) under Creative Commons Attribution 4.0 International.

SUPPLEMENTARY MATERIAL

# S1

## Effect of series length on ARFIMA and DFA outcomes (Hausdorff 1996, Tier 3)

*Philippe Terrier - HE-ARC Sante, Neuchatel, Switzerland*

# Supplementary Material — Effect of Series Length on ARFIMA and DFA Outcomes (Hausdorff 1996 Dataset, Unconstrained Walking)

Philippe Terrier — HE-ARC Santé, Neuchâtel, Switzerland.

This supplementary document can be read independently of the main manuscript. It extends the mixed-model analysis of the Hausdorff (1996) dataset to a third, longer segmentation tier (2000 strides) in order to evaluate the sensitivity of the ARFIMA/DFA outcomes to the length of the analysed time series.

## 1. Background and motivation

The main paper applies a family of eight nested time-series models, from white noise to the full ARFIMA(1,d,1), to stride-interval series from three publicly available datasets. Models are fitted by exact maximum likelihood, compared through the Schwarz (BIC) criterion, and averaged via Bayesian Model Averaging (BMA). The fractional differencing parameter *d* and the Detrended Fluctuation Analysis (DFA) scaling exponent *α* are the primary long-range correlation outcomes. The Schwarz weights W provide a continuous measure of evidence for each model.

The Hausdorff (1996) dataset (PhysioNet; 10 healthy adults, ≈ 1 hour of overground walking in six conditions) is the only dataset long enough to allow segmentation studies. In the main paper, two tiers are analysed: **T2** (L = 600 strides, multi-bout, primary analysis) and **T1** (L = 1200 strides, single bout per subject × condition). Both tiers include the three unconstrained conditions (slow, normal, fast) and the three metronomic conditions.

Because long-memory identification becomes easier as the sample size grows (the BIC penalty is dominated by the asymptotic log-likelihood gain), the question arose whether a longer segmentation would alter the picture obtained at L = 1200. A third tier, **T3_UNC** (L = 2000 strides, single bout), was added to address this point. Metronomic series are too short for this length (all are below 2000 strides), so T3_UNC contains only unconstrained bouts: 10 subjects × 3 speeds = 30 series. To keep the comparison clean, this supplementary analysis contrasts T3_UNC with the unconstrained subset of T1, i.e. the same 30 subject × speed combinations at two different lengths.

## 2. Methods

Stride-interval series from the PhysioNet UMWDB were truncated to the first 2000 strides of each recording after removing any negative or zero values (no transient cut-off). Each series was z-scored and then fitted with the eight-model family used in the main paper: white noise, AR(1), MA(1), ARMA(1,1), FI(d), ARFI(1,d,0), FIMA(0,d,1), and ARFIMA(1,d,1). Model fitting used the R package *arfima* (v1.8) with *dmean = FALSE* and *lmodel = "d"*. Random restarts followed the configuration of the main paper (numeach = c(5,5) for ARFI/FIMA/ARFIMA and c(1,1) for FI). The DFA exponent α was computed with polynomial order 1, 30 log-spaced box sizes between 16 and L/3, and least-squares fitting on the log-log plane.

For each outcome, a mixed-effects model was fitted separately at T1 and T3_UNC. To keep the T3_UNC estimates comparable with T1 while respecting the unconstrained-only design, the T3_UNC models used a speed-only fixed-effects formula (OUTCOME ~ speed_c + (1 | subject)), and the emmeans presented below are evaluated at speed_c = 0 (i.e. the normal-pace intercept). T1 emmeans are taken from the unconstrained rows of the main analysis (OUTCOME ~ pacing_c ×

speed_c + (1 | subject)), at speed_c = 0. Linear mixed models (LMMs, lme4 / lmerTest) were used for *d, DFA α, φ, θ*, and H_norm; beta generalised linear mixed models (glmmTMB, logit link) were used for the weights, which are bounded in (0, 1). Confidence intervals are reported at the 99% level, matching the main paper.

Two summary measures of the length effect are reported alongside the emmeans: (i) the absolute bias Δ = T3 − T1 on the original outcome scale, and (ii) the relative bias in percent, (T3 − T1) / T1 × 100, which quantifies the change independently of the units.

## 3. Results table

Table S1 reports the marginal means at both tiers with 99% confidence intervals, the absolute difference (bias), and the relative difference in percent. Rows are grouped by parameter family: long-range correlation primary outcomes, BMA-averaged ARMA correction, long-memory evidence (aggregate and per-model), and the normalised entropy of the weight distribution.

**Table S1.** ARFIMA and DFA outcomes at two series lengths in the unconstrained walking condition of the Hausdorff (1996) dataset. Estimated marginal means with 99% confidence intervals obtained from mixed-effects models (N = 10 subjects × 3 speeds per tier). Bias = T3 − T1; Rel. diff. = 100 × (T3 − T1) / T1.

| Outcome | Hausdorff (1996) — unconstrained walking<br>N = 10 subjects × 3 speeds | | Length effect<br>T3_UNC − T1 | |
|---|---|---|---|---|
| | T1 (L = 1200 strides) | T3_UNC (L = 2000 strides) | Bias (Δ) | Rel. diff. (%) |
| ***d_BMA***<br>(LMM) | 0.230<br>[0.091, 0.368] | 0.252<br>[0.138, 0.367] | +0.023 | +9.9% |
| ***DFA α***<br>(LMM) | 0.977<br>[0.899, 1.054] | 0.976<br>[0.924, 1.029] | +0.000 | −0.1% |
| ***φ_BMA***<br>(LMM) | 0.253<br>[0.058, 0.447] | 0.220<br>[−0.039, 0.478] | −0.033 | **−13.1%** |
| ***θ_BMA***<br>(LMM) | −0.205<br>[−0.340, −0.070] | −0.205<br>[−0.338, −0.072] | +0.000 | +0.1% |
| ***W_long***<br>(beta) | 0.740<br>[0.602, 0.842] | 0.836<br>[0.706, 0.915] | +0.096 | **+13.0%** |
| ***W_FI***<br>(beta) | 0.113<br>[0.060, 0.203] | 0.087<br>[0.038, 0.188] | −0.025 | **−22.6%** |
| ***W_ARFI***<br>(beta) | 0.393<br>[0.260, 0.545] | 0.386<br>[0.256, 0.536] | −0.007 | −1.8% |
| ***W_FIMA***<br>(beta) | 0.134<br>[0.075, 0.229] | 0.163<br>[0.087, 0.285] | +0.029 | **+21.7%** |
| ***W_ARFIMA***<br>(beta) | 0.296<br>[0.185, 0.437] | 0.310<br>[0.197, 0.452] | +0.014 | +4.8% |
| ***W_ARMA11***<br>(beta) | 0.232<br>[0.139, 0.361] | 0.164<br>[0.085, 0.294] | −0.068 | **−29.4%** |
| ***H_norm***<br>(LMM) | 0.320<br>[0.198, 0.443] | 0.335<br>[0.180, 0.490] | +0.014 | +4.5% |

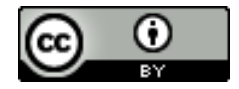

Outcomes: *d_BMA* = BMA-weighted fractional differencing parameter; *DFA α* = Detrended Fluctuation Analysis scaling exponent; *φ_BMA* and *θ_BMA* = BMA-weighted AR(1) and MA(1) coefficients (standard sign convention, $1 + \theta B$); *W_FI, W_ARFI, W_FIMA, W_ARFIMA* = Schwarz weights of the four long-memory models; *W_long* = their sum (evidence for any long-memory model); *W_ARMA11* = Schwarz weight of the short-memory ARMA(1,1) model; *H_norm* = Shannon entropy of the 8 Schwarz weights, normalised by log(8), so that 0 = a single model carries all the evidence and 1 = maximum uncertainty across the 8 models.

Model class: (LMM) = linear mixed model fitted with lme4/lmerTest, Kenward–Roger / Satterthwaite CIs; (beta) = beta GLMM fitted with glmmTMB (logit link), CIs back-transformed to the response scale. All models include a random intercept per subject ((1 | subject)). T1 fixed effects: pacing_c × speed_c; T3_UNC fixed effects: speed_c only (pacing is constant). Confidence level: 99%. Cells in the relative-difference column are shaded when the absolute relative change exceeds 10%.

## 4. Interpretation

**Stable parameters.** The three descriptive parameters of the long-range correlation structure — the BMA-averaged fractional differencing *d*, the DFA exponent α, and the BMA-averaged ARMA coefficients φ and θ — are remarkably stable between the two lengths. DFA α changes by less than 0.001 units (≈ 0.1%), θ_BMA by essentially zero, *d* by about +0.02 (+10%), and φ_BMA by roughly −0.03 (−13%). The 99% confidence intervals overlap broadly for all four outcomes. These point estimates are the values that would typically be reported for a given participant, and they do not depend sensibly on whether the series is 1200 or 2000 strides long.

**Reorganisation of the weights.** In contrast, the Schwarz weights — which encode how much evidence the data provide *for each candidate model* — change in a direction that is informative and consistent with classical BIC theory. The aggregate long-memory weight W_long rises from 0.74 to 0.84 (+13%), while the short-memory competitor W_ARMA11 falls from 0.23 to 0.16 (−29%). Within the long-memory family, evidence migrates from the simpler FI(d) model (−23%) towards the richer FIMA (+22%) and, to a lesser extent, ARFIMA (+5%) parameterisations, while ARFI remains essentially unchanged.

**Mechanism.** These shifts are expected. BIC penalises model complexity by an amount proportional to log(N); as N grows, the log-likelihood gain of the more flexible nested models (ARFIMA ⊃ FIMA ⊃ FI) dominates the penalty, and their BIC weight rises. The short-memory family (ARMA(1,1) in particular) is a poor parameterisation of a truly long-memory process: at short lengths BIC can still make it competitive through parsimony, but at longer lengths the likelihood gap becomes too wide to overcome.

**Practical take-away.** For the purpose of reporting a single estimate of long-range persistence per participant, the 1200-stride analysis of the main paper is adequate: *d_BMA* and DFA α are essentially invariant to the length. However, any statement about *which specific model is preferred* is contingent on the length of the series, and our results reproduce the textbook behaviour that longer series increasingly favour the full ARFIMA parameterisation over both its sub-models and the short-memory alternatives. The normalised entropy H_norm stays close to 0.33 at both tiers, confirming that the weight distribution remains moderately concentrated: BIC consistently picks a clear winner, even if the identity of that winner shifts with length.

## 5. Data and code availability

The raw stride-interval recordings come from the PhysioNet "Long-term Recordings of Gait Dynamics" collection (Hausdorff et al. 1996; Goldberger et al. 2000; doi:10.13026/C28679), redistributed under the Open Data Commons Attribution Licence v1.0.

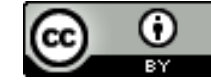

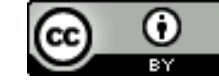

SUPPLEMENTARY MATERIAL

# S2

## Morris elementary-effects screening of the ARFIMA/DFA pipeline

*Philippe Terrier - HE-ARC Sante, Neuchatel, Switzerland*

# Supplementary Material — Morris Elementary Effects Screening of the ARFIMA/DFA Gait Analysis Pipeline

Philippe Terrier — HE-ARC Santé, Neuchâtel, Switzerland.

This supplementary document can be read independently of the main manuscript. It provides the complete results of the Morris Elementary Effects screening applied to the ARFIMA/DFA gait analysis pipeline. The screening identifies which pipeline parameters influence the scientific outputs (fractional differencing parameter *d*, long-memory classification, and DFA scaling exponent α) and which can be fixed at default values without affecting conclusions. The interpretation of these results and the resulting parameter choices are summarised in Section 3.5 of the main manuscript.

## 1. Screening design

Two separate Morris screenings were conducted on the Terrier (2016) stride time data, using the normal walking (N_ST) and auditory-cued walking (A_ST) conditions (36 participants × 500 strides each). The ARFIMA and DFA algorithms were screened independently because they share only preprocessing parameters; algorithm-specific parameters of one method have zero effect on the other's output.

**Table S2.1. Morris design specifications.**

| Setting | Value | Note |
|---|---|---|
| Trajectories (r) | 20 | Selected from 500 candidates |
| Levels (p) | 6 | Even, per Morris (1991) |
| Grid jump (Δ) | 3 (= p/2) | Standard recommendation |
| Scaling | TRUE | Required: different parameter ranges |
| Space-filling optimization | Campolongo et al. (2007) | 500 candidate trajectories |
| Dataset | Terrier (2016) | 36 participants × 500 strides |
| Conditions | N_ST, A_ST | Normal and auditory-cued walking |
| ARFIMA configurations | 160 | 20 × (7 + 1) |
| DFA configurations | 140 | 20 × (6 + 1) |

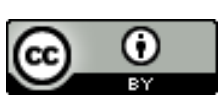

## 2. Parameter definitions and encoding

**Table S2.2a. ARFIMA screening parameters (k = 7).**

| ID | Parameter | Type | Levels | Values |
|---|---|---|---|---|
| P1 | Detrending | Ordinal | 3 | None, linear, quadratic |
| P2 | Outlier handling | Binary | 2 | None, ±3SD winsorization |
| A1 | numeach | Ordinal | 4 | (1,1), (2,2), (3,3), (5,5) |
| A2 | dmean | Binary | 2 | FALSE, TRUE |
| A3 | lmodel | Binary | 2 | d (fractional diff.), g (fGn) |
| A4 | Information criterion | Binary | 2 | BIC, AIC |
| A5 | d aggregation | Categorical | 3 | BMA, best-model, naïve average |

**Table S2.2b. DFA screening parameters (k = 6).**

| ID | Parameter | Type | Levels | Values |
|---|---|---|---|---|
| P1 | Detrending | Ordinal | 3 | None, linear, quadratic |
| P2 | Outlier handling | Binary | 2 | None, ±3SD winsorization |
| D1 | DFA polynomial order | Ordinal | 3 | DFA-1, DFA-2, DFA-3 |
| D2 | Minimum box size (n_min) | Ordinal | 4 | 4, 8, 10, 16 |
| D3 | Box size spacing | Categorical | 3 | Dense-additive, log-moderate, log-sparse |
| D4 | Regression method | Binary | 2 | OLS, robust (Theil-Sen) |

*Note. Parameters P1 and P2 (preprocessing) appear in both screenings. Fixed DFA settings not screened: n_max = N/4, non-overlapping boxes.*

**Table S2.2c. Parameters excluded from screening, with rationale.**

| Parameter | Rationale for exclusion |
|---|---|
| Normalization (z-score) | Invariant: z-scoring is a linear transform that does not affect d or α. |
| Series length truncation | Effect dominated by well-established theoretical results (Braun 2010). All Terrier series are 500 strides. |
| Model space (8 vs. 12 models) | Higher-order short-memory models (AR(2), ARMA(2,1)) add estimation complexity at T = 500 without improving discrimination. |
| Transient removal | Minimal transient effects in the Terrier protocol (warm-up before recording; continuous walking). |
| DFA n_max (N/3, N/4, N/5) | Range too narrow at T = 500 (100–167 strides). Fixed at N/4 (standard default). |
| DFA overlap (yes/no) | Non-overlapping is the universal standard in gait research. |

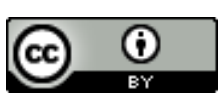

## 3. Morris summary statistics ($\mu$, $\mu^*$, $\sigma$)

Tables S2.3a–e report the Morris summary statistics for each parameter and output metric. Parameters are sorted by decreasing $\mu^*$ (mean absolute elementary effect). The classification follows standard criteria: parameters with $\mu^*$ below 10% of the maximum $\mu^*$ are labelled “non-influential”; among the remaining parameters, those with $\sigma/\mu^* < 1$ are labelled “additive” (consistent effect direction) and those with $\sigma/\mu^* \geq 1$ are labelled “interactive” (effect depends on other parameter values).

**Table S2.3a. ARFIMA screening — Output: mean d (N_ST, normal walking).**

| Parameter | $\mu$ | $\mu^*$ | $\sigma$ | $\sigma/\mu^*$ | Category | Interpretation |
|---|---|---|---|---|---|---|
| **A5_d_agg** | 0.122 | **0.155** | 0.170 | 1.098 | *Interactive* | Effect varies by context |
| **A1_numeach** | -0.130 | **0.130** | 0.085 | 0.653 | *Additive* | More restarts reduce d (better optima found) |
| **P1_detrend** | -0.119 | **0.119** | 0.095 | 0.798 | *Additive* | Detrending reduces d (removes low-freq. content) |
| **A4_criterion** | -0.106 | **0.106** | 0.106 | 1.002 | *Interactive* | BIC yields lower d than AIC |
| **A3_lmodel** | 0.029 | **0.105** | 0.136 | 1.304 | *Interactive* | Effect varies by context |
| **A2_dmean** | 0.038 | **0.039** | 0.047 | 1.212 | *Interactive* | Marginal |
| P2_outlier | -0.004 | 0.013 | 0.016 | 1.251 | *Non-influential* | Negligible |

**Table S2.3b. ARFIMA screening — Output: mean d (A_ST, auditory-cued walking).**

| Parameter | $\mu$ | $\mu^*$ | $\sigma$ | $\sigma/\mu^*$ | Category | Interpretation |
|---|---|---|---|---|---|---|
| **A3_lmodel** | 0.525 | **0.525** | 0.156 | 0.297 | *Additive* | Switching d → g shifts d by ≈ +0.5 |
| **A5_d_agg** | 0.186 | **0.212** | 0.231 | 1.089 | *Interactive* | Effect varies by context |
| **A1_numeach** | 0.073 | **0.116** | 0.147 | 1.265 | *Interactive* | Effect varies by context |
| **A2_dmean** | -0.060 | **0.068** | 0.056 | 0.817 | *Additive* | dmean = TRUE pushes d more negative |
| **A4_criterion** | -0.063 | **0.065** | 0.061 | 0.942 | *Additive* | BIC yields slightly more negative d |
| P1_detrend | -0.016 | 0.034 | 0.043 | 1.264 | *Non-influential* | Negligible effect on anti-persistent d |
| P2_outlier | 0.001 | 0.027 | 0.037 | 1.393 | *Non-influential* | Negligible |

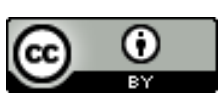

**Table S2.3c. ARFIMA screening — Output: % long-memory classification (N_ST).**

| Parameter | μ | μ* | σ | σ/μ* | Category | Interpretation |
|---|---|---|---|---|---|---|
| **A3_lmodel** | -15.509 | **16.435** | 13.871 | 0.844 | *Additive* | fGn reduces long-memory classification |
| **A1_numeach** | 13.657 | **13.657** | 13.306 | 0.974 | *Additive* | More restarts increase long-memory classification |
| **A4_criterion** | 5.093 | **13.426** | 15.818 | 1.178 | *Interactive* | Direction varies by context |
| **P1_detrend** | 0.231 | **9.954** | 14.048 | 1.411 | *Interactive* | Direction varies by context |
| **A2_dmean** | -5.324 | **5.787** | 8.265 | 1.428 | *Interactive* | Direction varies by context |
| **P2_outlier** | 3.241 | **5.093** | 5.004 | 0.983 | *Additive* | Marginal increase in long-memory classification |
| A5_d_agg | 0.000 | 0.000 | 0.000 | — | *Non-influential* | No effect (classification is model-based) |

*Note. μ and μ* are in percentage-point units. A5 (d aggregation) has zero effect because long-memory classification depends on the best model, not on how d is aggregated across models.*

**Table S2.3d. DFA screening — Output: mean α (N_ST, normal walking).**

| Parameter | μ | μ* | σ | σ/μ* | Category | Interpretation |
|---|---|---|---|---|---|---|
| **D1_poly_order** | -0.089 | **0.095** | 0.064 | 0.677 | *Additive* | Higher order reduces α |
| **D3_spacing** | -0.038 | **0.067** | 0.068 | 1.012 | *Interactive* | Effect varies by context |
| **D2_n_min** | -0.004 | **0.065** | 0.092 | 1.427 | *Interactive* | Effect varies by context |
| **D4_reg_method** | -0.007 | **0.025** | 0.032 | 1.287 | *Interactive* | Marginal |
| P1_detrend | -0.005 | 0.005 | 0.009 | 1.747 | *Non-influential* | Negligible |
| P2_outlier | -0.001 | 0.001 | 0.002 | 1.131 | *Non-influential* | Negligible |

**Table S2.3e. DFA screening — Output: mean α (A_ST, auditory-cued walking).**

| Parameter | μ | μ* | σ | σ/μ* | Category | Interpretation |
|---|---|---|---|---|---|---|
| **D1_poly_order** | 0.162 | **0.162** | 0.059 | 0.367 | *Additive* | Higher order increases α |
| **D3_spacing** | 0.158 | **0.158** | 0.092 | 0.582 | *Additive* | Log-sparser spacing increases α |
| **D2_n_min** | -0.151 | **0.151** | 0.107 | 0.708 | *Additive* | Larger n_min reduces α |
| **D4_reg_method** | -0.063 | **0.063** | 0.059 | 0.943 | *Additive* | Theil-Sen reduces α |
| P2_outlier | -0.016 | 0.016 | 0.006 | 0.357 | *Non-influential* | Negligible |
| P1_detrend | 0.000 | 0.000 | 0.000 | 1.568 | *Non-influential* | Negligible |

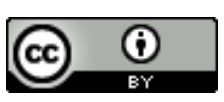

## 4. Elementary effects matrices

Tables S2.4a–e present the individual elementary effects from each of the r = 20 trajectories. Each row corresponds to one trajectory; each column to one parameter. These values are the raw data from which the summary statistics in Section S2.3 were computed (μ = mean, μ* = mean of absolute values, σ = standard deviation).

**Table S2.4a. Elementary effects — mean d (N_ST).**

| Traj. | P1 | P2 | A1 | A2 | A3 | A4 | A5 |
|---|---|---|---|---|---|---|---|
| # | *Detrend.* | *Outlier* | *numeach* | *dmean* | *Imodel* | *Criterion* | *d agg.* |
| 1 | -0.0542 | 0.0010 | -0.0649 | -0.0015 | -0.0549 | -0.0353 | -0.0110 |
| 2 | -0.1031 | -0.0169 | -0.0661 | 0.0000 | -0.0230 | -0.1242 | 0.3187 |
| 3 | -0.0503 | -0.0118 | -0.2366 | 0.0000 | 0.3176 | 0.0000 | 0.3293 |
| 4 | -0.1899 | -0.0116 | -0.2360 | -0.0056 | -0.0266 | -0.3157 | 0.0047 |
| 5 | -0.1566 | 0.0267 | -0.0884 | 0.0442 | 0.0992 | -0.1779 | -0.0012 |
| 6 | -0.0314 | 0.0103 | -0.0654 | 0.0009 | -0.1040 | -0.0261 | 0.2493 |
| 7 | -0.0632 | -0.0364 | -0.2120 | 0.0999 | -0.1034 | 0.0000 | 0.2343 |
| 8 | -0.2975 | -0.0080 | -0.1497 | 0.0198 | 0.2329 | -0.0834 | 0.2933 |
| 9 | -0.0423 | -0.0186 | -0.1623 | 0.1028 | -0.0151 | -0.1781 | 0.0499 |
| 10 | -0.1279 | -0.0024 | 0.0000 | 0.0999 | -0.0817 | -0.0469 | 0.0889 |
| 11 | -0.0504 | -0.0183 | -0.1163 | 0.0022 | -0.0345 | -0.2155 | -0.0259 |
| 12 | -0.0577 | 0.0178 | -0.0004 | 0.0795 | -0.1193 | -0.0623 | 0.1606 |
| 13 | -0.1974 | 0.0128 | -0.2110 | 0.0017 | -0.0279 | -0.1242 | -0.0846 |
| 14 | -0.0519 | 0.0014 | -0.1506 | 0.0014 | -0.1214 | -0.1389 | 0.0748 |
| 15 | -0.0883 | 0.0058 | -0.0286 | 0.1039 | -0.0483 | -0.0162 | 0.2902 |
| 16 | -0.0494 | 0.0032 | -0.2793 | 0.0451 | 0.2687 | -0.0241 | -0.1590 |
| 17 | -0.1541 | -0.0281 | -0.0427 | 0.1343 | 0.0775 | 0.0000 | 0.0789 |
| 18 | -0.0903 | 0.0118 | -0.1080 | 0.0212 | 0.0669 | -0.3513 | 0.0863 |
| 19 | -0.4035 | -0.0064 | -0.1562 | 0.0100 | 0.2342 | 0.0000 | 0.5081 |
| 20 | -0.1274 | -0.0023 | -0.2325 | 0.0001 | 0.0350 | -0.1970 | -0.0428 |

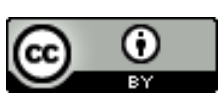

**Table S2.4b. Elementary effects — mean d (A_ST).**

| Traj. | P1 | P2 | A1 | A2 | A3 | A4 | A5 |
|---|---|---|---|---|---|---|---|
| # | *Detrend.* | *Outlier* | *numeach* | *dmean* | *lmodel* | *Criterion* | *d agg.* |
| 1 | 0.0456 | -0.0942 | 0.2561 | -0.1006 | 0.6558 | -0.0345 | 0.0152 |
| 2 | 0.0068 | 0.0159 | -0.0251 | -0.1042 | 0.3663 | -0.1395 | 0.5469 |
| 3 | -0.0103 | 0.0029 | 0.1358 | 0.0828 | 0.7834 | 0.0000 | 0.2205 |
| 4 | -0.1066 | 0.0113 | -0.0297 | -0.1192 | 0.3602 | -0.0965 | 0.4009 |
| 5 | -0.0602 | 0.0157 | 0.0000 | -0.0059 | 0.5291 | -0.0225 | -0.0361 |
| 6 | -0.0134 | 0.0051 | -0.0278 | -0.0323 | 0.3720 | -0.1638 | 0.5704 |
| 7 | -0.0011 | 0.0135 | 0.3599 | -0.0988 | 0.4100 | 0.0000 | 0.0546 |
| 8 | -0.0752 | -0.0157 | -0.0136 | -0.0127 | 0.4820 | -0.1286 | 0.3114 |
| 9 | -0.0052 | 0.0128 | -0.0302 | -0.0912 | 0.4051 | -0.0872 | -0.0583 |
| 10 | 0.0105 | -0.0332 | -0.1201 | -0.0988 | 0.7060 | -0.1578 | -0.0581 |
| 11 | -0.0332 | -0.0104 | -0.0703 | -0.0661 | 0.3585 | -0.0760 | -0.0171 |
| 12 | -0.0222 | -0.0651 | 0.2737 | -0.0326 | 0.3609 | -0.0280 | 0.3496 |
| 13 | -0.0673 | 0.0321 | 0.1122 | -0.1301 | 0.4274 | -0.1395 | -0.0161 |
| 14 | -0.0054 | 0.0032 | 0.2776 | -0.0948 | 0.4512 | 0.0091 | 0.0032 |
| 15 | 0.0115 | -0.0341 | -0.0364 | -0.0965 | 0.5845 | -0.1236 | 0.4544 |
| 16 | -0.0220 | 0.0353 | -0.0448 | 0.0003 | 0.4290 | -0.0371 | -0.0501 |
| 17 | -0.0755 | 0.0754 | 0.1937 | -0.0319 | 0.7235 | 0.0000 | 0.4819 |
| 18 | 0.0473 | 0.0203 | 0.0117 | -0.0346 | 0.6361 | -0.0489 | 0.4187 |
| 19 | 0.0563 | 0.0043 | -0.0319 | -0.0012 | 0.6153 | 0.0000 | 0.1534 |
| 20 | -0.0104 | 0.0316 | 0.2736 | -0.1295 | 0.8386 | 0.0055 | -0.0258 |

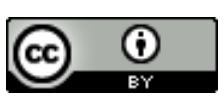

**Table S2.4c. Elementary effects — % long-memory classification (N_ST).**

| Traj. | P1 | P2 | A1 | A2 | A3 | A4 | A5 |
|---|---|---|---|---|---|---|---|
| # | *Detrend.* | *Outlier* | *numeach* | *dmean* | *lmodel* | *Criterion* | *d agg.* |
| 1 | 0.0000 | 4.63 | 9.26 | 0.0000 | -23.15 | 13.89 | 0.0000 |
| 2 | -4.63 | 4.63 | 13.89 | 0.0000 | -13.89 | -27.78 | 0.0000 |
| 3 | 0.0000 | 9.26 | 37.04 | 0.0000 | -13.89 | -4.63 | 0.0000 |
| 4 | 18.52 | 4.63 | 9.26 | 0.0000 | 4.63 | 4.63 | 0.0000 |
| 5 | -4.63 | 9.26 | 0.0000 | 0.0000 | -23.15 | 23.15 | 0.0000 |
| 6 | -32.41 | -9.26 | 9.26 | -4.63 | -18.52 | -4.63 | 0.0000 |
| 7 | 4.63 | 4.63 | 32.41 | -18.52 | -18.52 | 18.52 | 0.0000 |
| 8 | -4.63 | 0.0000 | 32.41 | -4.63 | -41.67 | 13.89 | 0.0000 |
| 9 | -9.26 | 9.26 | 4.63 | -18.52 | 0.0000 | 0.0000 | 0.0000 |
| 10 | 4.63 | 0.0000 | 0.0000 | -18.52 | 0.0000 | 0.0000 | 0.0000 |
| 11 | -4.63 | -4.63 | 0.0000 | 0.0000 | -23.15 | -4.63 | 0.0000 |
| 12 | -9.26 | 4.63 | 0.0000 | 4.63 | -23.15 | 23.15 | 0.0000 |
| 13 | 0.0000 | 4.63 | 27.78 | 0.0000 | 0.0000 | -27.78 | 0.0000 |
| 14 | 18.52 | 9.26 | 32.41 | -4.63 | -4.63 | 27.78 | 0.0000 |
| 15 | -23.15 | 0.0000 | 0.0000 | -23.15 | -23.15 | -4.63 | 0.0000 |
| 16 | -4.63 | 4.63 | 18.52 | 0.0000 | -46.30 | -9.26 | 0.0000 |
| 17 | 18.52 | 4.63 | 4.63 | -13.89 | -18.52 | 18.52 | 0.0000 |
| 18 | 4.63 | 4.63 | 0.0000 | -4.63 | -9.26 | 13.89 | 0.0000 |
| 19 | 27.78 | 4.63 | 13.89 | 0.0000 | -18.52 | 9.26 | 0.0000 |
| 20 | 4.63 | -4.63 | 27.78 | 0.0000 | 4.63 | 18.52 | 0.0000 |

*Note. Values are in percentage-point units.*

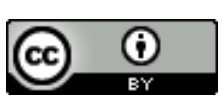

**Table S2.4d. Elementary effects — mean α (N_ST).**

| Traj. | P1 | P2 | D1 | D2 | D3 | D4 |
|---|---|---|---|---|---|---|
| # | *Detrend.* | *Outlier* | *Poly. order* | *n_min* | *Spacing* | *Regr. method* |
| 1 | 0.0000 | -0.0029 | -0.0598 | 0.0796 | -0.0856 | 0.0293 |
| 2 | -0.0088 | -0.0040 | -0.1349 | 0.0607 | -0.0547 | -0.0106 |
| 3 | -0.0028 | 0.0004 | -0.0763 | 0.0135 | -0.0886 | 0.0195 |
| 4 | -0.0139 | 0.0004 | -0.0884 | 0.0628 | -0.0499 | 0.0149 |
| 5 | 0.0003 | 0.0000 | -0.0946 | 0.0422 | -0.0114 | -0.0041 |
| 6 | 0.0005 | -0.0002 | 0.0601 | 0.0376 | -0.0725 | -0.0134 |
| 7 | 0.0000 | -0.0025 | -0.0472 | -0.2262 | 0.1255 | -0.0199 |
| 8 | -0.0105 | -0.0003 | -0.1438 | -0.0477 | -0.0267 | -0.0042 |
| 9 | 0.0000 | 0.0008 | -0.2148 | -0.0167 | -0.0804 | 0.0000 |
| 10 | -0.0053 | 0.0000 | -0.0964 | 0.0954 | -0.0657 | -0.0163 |
| 11 | -0.0382 | -0.0001 | -0.0540 | 0.0679 | -0.0553 | 0.0252 |
| 12 | -0.0112 | -0.0025 | -0.0879 | -0.0047 | -0.1713 | -0.0208 |
| 13 | 0.0008 | -0.0033 | -0.0403 | -0.2262 | 0.0828 | -0.0919 |
| 14 | 0.0002 | -0.0027 | -0.0275 | -0.0191 | -0.0568 | -0.0167 |
| 15 | 0.0004 | -0.0007 | -0.1970 | 0.0628 | -0.0368 | 0.0317 |
| 16 | 0.0000 | -0.0004 | -0.0811 | 0.0053 | -0.0356 | -0.0199 |
| 17 | -0.0151 | 0.0003 | -0.1858 | 0.0614 | -0.0295 | 0.0308 |
| 18 | 0.0000 | -0.0017 | -0.0819 | -0.0190 | -0.0864 | -0.0486 |
| 19 | 0.0000 | -0.0035 | -0.0276 | -0.1304 | 0.0851 | -0.0549 |
| 20 | 0.0002 | -0.0004 | -0.0983 | 0.0118 | -0.0386 | 0.0322 |

**Table S2.4e. Elementary effects — mean α (A_ST).**

| Traj. | P1 | P2 | D1 | D2 | D3 | D4 |
|---|---|---|---|---|---|---|
| # | *Detrend.* | *Outlier* | *Poly. order* | *n_min* | *Spacing* | *Regr. method* |
| 1 | 0.0000 | -0.0115 | 0.1205 | -0.1062 | 0.1707 | -0.1383 |
| 2 | -0.0008 | -0.0121 | 0.1602 | -0.2582 | 0.1938 | -0.0118 |
| 3 | -0.0005 | -0.0196 | 0.1895 | -0.0265 | 0.1757 | -0.0638 |
| 4 | 0.0004 | -0.0247 | 0.1420 | -0.0858 | 0.0509 | -0.0291 |
| 5 | -0.0001 | -0.0169 | 0.0994 | -0.1216 | 0.2224 | -0.0179 |
| 6 | 0.0000 | -0.0085 | 0.2702 | -0.1250 | 0.2719 | -0.0188 |
| 7 | 0.0000 | -0.0132 | 0.1376 | -0.3596 | 0.2898 | -0.0140 |
| 8 | -0.0005 | -0.0252 | 0.1930 | -0.1974 | 0.0974 | -0.2394 |
| 9 | 0.0000 | -0.0230 | 0.2937 | -0.1789 | 0.0905 | 0.0000 |
| 10 | -0.0004 | -0.0169 | 0.1606 | -0.1255 | 0.1306 | -0.0203 |

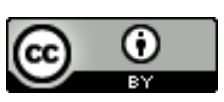

| | | | | | | |
|---|---|---|---|---|---|---|
| 11 | -0.0007 | -0.0150 | 0.1374 | -0.0818 | 0.1010 | -0.0792 |
| 12 | 0.0002 | -0.0132 | 0.1944 | -0.0982 | 0.2467 | -0.0128 |
| 13 | -0.0001 | -0.0092 | 0.1043 | -0.3596 | 0.2892 | -0.1111 |
| 14 | 0.0000 | -0.0121 | 0.1606 | -0.1284 | 0.1839 | -0.0202 |
| 15 | 0.0001 | -0.0122 | 0.2251 | -0.0858 | 0.0426 | -0.0843 |
| 16 | 0.0000 | -0.0116 | 0.1130 | -0.0373 | 0.0589 | -0.0233 |
| 17 | 0.0001 | -0.0191 | 0.2378 | -0.1597 | 0.0376 | -0.1085 |
| 18 | 0.0000 | -0.0093 | 0.0860 | -0.1102 | 0.1441 | -0.1105 |
| 19 | 0.0000 | -0.0121 | 0.1013 | -0.3601 | 0.3142 | -0.0703 |
| 20 | 0.0002 | -0.0256 | 0.1089 | -0.0084 | 0.0466 | -0.0793 |

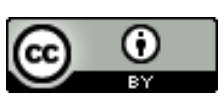

## 5. Pipeline output ranges

Table S2.5 summarises the range of each output metric across all pipeline configurations tested during the Morris screening. These ranges quantify the total sensitivity of the scientific conclusions to the set of defensible analysis choices.

**Table S2.5. Output summary statistics across all Morris configurations.**

| Output metric | N configs | Min | Max | Mean | SD | Screening |
|---|---|---|---|---|---|---|
| mean d (N_ST) | 160 | −0.356 | 0.315 | 0.088 | 0.121 | ARFIMA |
| mean d (A_ST) | 160 | −0.721 | −0.150 | −0.413 | 0.179 | ARFIMA |
| % long-memory (N_ST) | 160 | 33.3 | 72.2 | 52.4 | 8.9 | ARFIMA |
| mean $\alpha$ (N_ST) | 140 | 0.740 | 0.942 | 0.825 | 0.055 | DFA |
| mean $\alpha$ (A_ST) | 140 | 0.283 | 0.800 | 0.467 | 0.129 | DFA |

*Note. Each configuration represents one unique combination of parameter values from the Morris design matrix. Output values are group-level summaries across 36 participants per condition. % long-memory is the percentage of participants for whom the best ARFIMA model (by BIC) includes a fractional differencing parameter.*

## 6. Recommended parameter settings

Table S2.6 lists the parameter settings adopted for all subsequent analyses, with the screening-based rationale for each choice.

**Table S2.6. Final pipeline parameter settings.**

| ID | Parameter | Setting | Screening-based rationale |
|---|---|---|---|
| P1 | Detrending | None | Non-influential for A_ST; removing low-freq. content risks d underestimation in N_ST |
| P2 | Outlier handling | None | Non-influential across all outputs |
| A1 | numeach | c(5,5)† | Additive: more restarts consistently improve optimization quality ($\mu^*$ = 13.7 pp on classification) |
| A2 | dmean | FALSE | Additive on A_ST ($\mu^*$ = 0.068); FALSE is correct for z-scored data (Veenstra, 2013) |
| A3 | lmodel | d (fd) | Dominant parameter for A_ST ($\mu^*$ = 0.525); fd is the standard ARFIMA parameterization |
| A4 | Information criterion | BIC | Interactive; BIC's stronger complexity penalty is more conservative |
| A5 | d aggregation | BMA | Interactive for d; BMA is theoretically principled (Burnham & Anderson, 2002) |
| D1 | DFA polynomial order | DFA-1 | Most influential DFA parameter; DFA-1 is the gait literature standard |
| D2 | DFA n_min | 16 | Additive for A_ST ($\mu^*$ = 0.151); 16 is the recommended minimum (Damouras et al., 2010) |
| D3 | DFA box spacing | Log-geometric (30 pts) | Additive for A_ST ($\mu^*$ = 0.158); follows Almurad & Delignières (2016) |
| D4 | DFA regression | OLS | Least influential DFA parameter; OLS is standard |

*† numeach = c(5,5) for ARFI, FIMA, and ARFIMA models; c(1,0) for short-memory models (no multi-start needed); c(1,1) for FI (arfima package constraint).*

## 7. Data and code availability

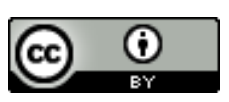

The Morris screening was performed on the Terrier (2016) stride time series (conditions N_ST and A_ST), redistributed under the Creative Commons Attribution 4.0 Licence. The screening R code, input data, configuration files, raw Morris outputs (trajectory-level elementary effects, μ/μ*/σ summary statistics), and the sensitivity figures are openly available as part of the reproduction archive deposited on Zenodo (https://doi.org/10.5281/zenodo.19676064), alongside the main pipeline and the mixed-effects analyses.

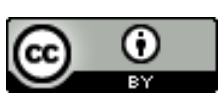